\documentclass[iop]{emulateapj}
\usepackage{natbib}
\usepackage{mathptmx} %
\usepackage[T1]{fontenc}
\usepackage{graphicx}

\def\kms {{\mathrm{km}\,\mathrm{s}^{-1}}}

\def\CaII{\ion{Ca}{2}}
\def\CaH{\CaII\, H}

\def\MgII{\ion{Mg}{2}}
\def\MgIIk{\ion{Mg}{2}\,k}
\def\MgIIh{\ion{Mg}{2}\,h}
\def\hk{{h\&k}}
\def\MgIIhk{\MgII\, \hk}

\def\kthree{\mbox{k$_3$}}    
\def\hthree{\mbox{h$_3$}}
\def\ktwo{\mbox{k$_2$}}
\def\htwo{\mbox{h$_2$}}

\def\ktwoV{\mbox{k$_{2V}$}}
\def\ktwoR{\mbox{k$_{2R}$}}

\def\htwoV{\mbox{h$_{2V}$}}
\def\htwoR{\mbox{h$_{2R}$}}
\def\honeV{\mbox{h$_{1V}$}}
\def\honeR{\mbox{h$_{1R}$}}

\def\ktwor{\mbox{k$_{2R}$}}

\def\RH{{\it RH}}
\def\bifrost{{\it Bifrost}}

\usepackage{mathptmx} 
\usepackage[colorlinks=true,urlcolor=blue,citecolor=blue,linkcolor=blue]{hyperref}

\defcitealias{Leenaarts:Mg1}{Paper I}
\defcitealias{Leenaarts:Mg2}{Paper II}

\shorttitle{The formation of \emph{IRIS} diagnostics. III.}
\shortauthors{Pereira et al.}

\begin{document}

\title{The formation of IRIS diagnostics. \\
III. Near-Ultraviolet Spectra and Images}
  
   \author{T. M. D. Pereira$^{1, 2, 3}$}\email{tiago.pereira@astro.uio.no}
   \author{J.~Leenaarts$^{1}$}%
   \author{B. De Pontieu$^{1, 3}$}%
   \author{M. Carlsson$^{1}$}%
   \author{H. Uitenbroek$^{4}$}%

\affil{$^1$ Institute of
  Theoretical Astrophysics, University of Oslo, P.O. Box 1029
  Blindern, N--0315 Oslo, Norway}
\affil{$^{2}$NASA Ames Research Center, Moffett Field, CA 94035, USA}
\affil{$^{3}$Lockheed Martin Solar and Astrophysics Laboratory, 3251 Hanover Street, Org. A021S, Bldg. 252, Palo Alto, CA 94304, USA}
\affil{$^4$NSO/Sacramento Peak P.O. Box 62
         Sunspot, NM 88349--0062 USA}

   \date{Received; accepted}

   \begin{abstract}

The \MgIIhk\ lines are the prime chromospheric diagnostics of NASA's Interface Region Imaging Spectrograph (IRIS).
In the previous papers of this series we used a realistic three-dimensional radiative magnetohydrodynamics model to calculate the \hk\ lines in detail and investigated how their spectral features relate to the underlying atmosphere.
 In this work we employ the same approach to investigate how the \hk\ diagnostics fare when taking into account the finite resolution of IRIS and different noise levels. In addition, we investigate the diagnostic potential of several other photospheric lines and near-continuum regions present in the near-ultraviolet (NUV) window of IRIS and study the formation of the NUV slit-jaw images.
We find that the instrumental resolution of IRIS has a small effect on the quality of the \hk\ diagnostics; the relations between the spectral features and atmospheric properties are mostly unchanged. The peak separation is the most affected diagnostic, but mainly due to limitations of the simulation. The effects of noise start to be noticeable at a signal-to-noise ratio (S/N)of 20, but we show that with noise filtering one can obtain reliable diagnostics at least down to a S/N of 5.
The many photospheric lines present in the NUV window provide velocity information for at least eight distinct photospheric heights. Using line-free regions in the \hk\ far wings we derive good estimates of photospheric temperature for at least three heights. Both of these diagnostics, in particular the latter, can be obtained even at S/Ns as low as 5.

\end{abstract}


\section{Introduction}                          \label{sec:introduction}
The \MgIIhk\ resonance lines are promising tools to study the solar chromosphere. Their location in the UV spectrum precludes ground-based observations, thus they have been observed with rocket \citep[e.g.][]{Bates:1969, Kohl:1976, Allen:1978, Morrill:2008, West:2011}, balloon \citep[e.g.][]{Lemaire:1969, Samain:1985, Staath:1995}, or satellite \citep[e.g.][]{Doschek:1977, Bonnet:1978, Woodgate:1980} experiments. The relatively high abundance of Mg together with their high oscillator strengths place these lines among the strongest in the solar UV spectrum; their radiation samples regions from the upper photosphere to the upper chromosphere \citep{Milkey:1974, Uitenbroek:1997}. When observed at disk-centre, the mean spectra from both lines show strong absorption with double-peaked emission cores, which reflect the atmospheric structure in their formation region and are markedly different from mean spectra of other widely used chromospheric lines such as \CaH~\&~K (formed deeper in the solar atmosphere). Modeling the \MgIIhk\ radiation is a complicated affair. Being very strong resonance lines, they suffer from partial frequency redistribution \citep[PRD, see][]{Milkey:1974} and given their chromospheric nature, the approximation of Local Thermodynamic Equilibrium (LTE) is not valid. While earlier studies relied on one-dimensional model atmospheres, the \MgIIhk\ modeling efforts have been recently pushed forward by \citet[hereafter \citetalias{Leenaarts:Mg1}]{Leenaarts:Mg1} and \citet[hereafter \citetalias{Leenaarts:Mg2}]{Leenaarts:Mg2}, who made use of a realistic 3D radiative MHD simulation of the solar atmosphere and detailed radiative transfer calculations.

The IRIS mission is an ambitious space observatory that seeks to understand how internal convective flows energize the solar atmosphere. Its main science goals are to find out (1) which types of non-thermal energy dominate in the chromosphere and beyond, (2) how does the chromosphere regulate the mass and energy supply to the corona and heliosphere, and (3) how do magnetic flux and matter rise through the lower atmosphere and what role does flux emergence play in flares and mass ejections. To address these questions, IRIS has a spectrograph (De Pontieu et al. 2013, in preparation) that observes in two ultraviolet bands (far-ultraviolet (FUV): $133.2-140.6$~nm; near-ultraviolet (NUV): $278.3-283.4$~nm) covering several lines that together sample a wide range of formation heights from the photosphere, chromosphere, transition region, and up to the corona. The main objective of the NUV window is to observe the \MgIIhk\ lines. The many earlier observations of these lines were either of low spatial or low spectral resolution. IRIS will provide a dramatically improved view of the chromosphere by observing the \hk\ lines at high cadence (1~s) and high spectral resolution ($\approx$6~pm) at a spatial resolution of $0\farcs4$.

\begin{figure*}
\begin{center}
\includegraphics[scale=0.8]{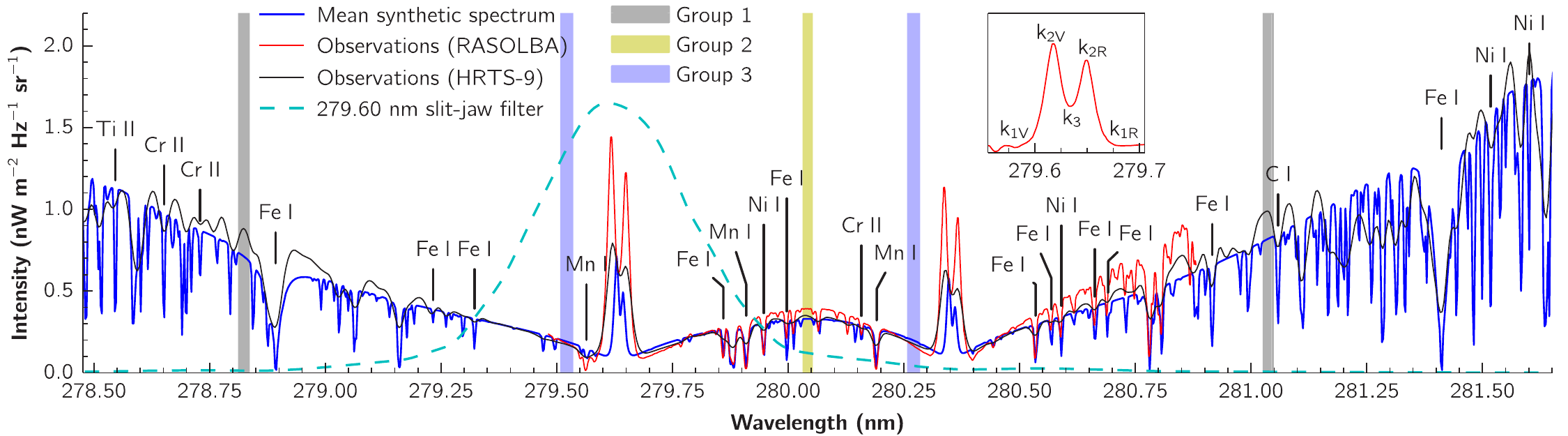}
\end{center}
\caption{Spatially and temporally averaged synthetic spectrum, compared with observations from RASOLBA \citep{Staath:1995} and HRTS-9 \citep{Morrill:2008}. The transmission function for the IRIS 279.60~nm slit-jaw filter is also shown (in arbitrary units). Also noted are the locations of the photospheric lines used in Sect.~\ref{sec:ph_vel}, and the quasi-continuum groups used in Sect.~\ref{sec:ph_temp}. The inset shows the location of the k$_{1}$, k$_{2}$, and k$_{3}$ spectral features.\label{fig:mspec_lines}}
\end{figure*}

The \MgIIhk\ lines are the most important chromospheric diagnostics in the NUV window of IRIS. Their formation was discussed in detail by \citetalias{Leenaarts:Mg1} and \citetalias{Leenaarts:Mg2}. In \citetalias{Leenaarts:Mg1}, a simplified Mg model atom for radiative transfer calculations was derived and the general formation properties of the \hk\ lines were discussed. Building on that, \citetalias{Leenaarts:Mg2} made use of a realistic solar model atmosphere and investigated how the \hk\ spectra relate to the atmosphere's thermodynamic conditions. In the present paper, we study the diagnostic potential contained in the IRIS NUV spectra and slit-jaw images, taking into account the finite resolution of the instrument and the impact of noise. We revisit the \MgII\ diagnostics as seen from IRIS and we also investigate the potential of the many photospheric lines in the NUV window of IRIS (mostly blended in the wings of the \hk\ lines). Our methodology follows closely that of \citetalias{Leenaarts:Mg2}: we make use of a realistic MHD simulation of the solar atmosphere, for which we calculate the detailed spectrum in the NUV window. The simulation employed does not reproduce some mean properties of the observed spectrum, see Figure~\ref{fig:mspec_lines}. In particular, the mean synthetic \MgII\ spectrum shows less emission than observed and the peak separation is too small. Our forward modelling gives a mapping between physical conditions in the model and the synthetic spectrum and this is valid even if the mean spectrum from the model does not match the observed mean spectrum. Applying this deduced mapping to the observations will tell us in what way the numerical simulations are inadequate in describing the Sun.

The outline of this paper is as follows. In Section~\ref{sec:models}, we describe the atmosphere model used and how the synthetic spectra were calculated. In Section~\ref{sec:mg}, we study the impact of the IRIS resolution on the derived relations between spectral features of \MgIIhk\ and physical conditions in the solar atmosphere and in Section~\ref{sec:phot} we study the diagnostic potential of several photospheric lines. In Section~\ref{sec:slit-jaw}, we study the formation properties of the NUV slit-jaw images. We conclude with a discussion in Section~\ref{sec:discussion}.

\section{Models and synthetic data}             \label{sec:models}

\subsection{Atmosphere Model}

To study the line formation in the NUV window of IRIS, we make use of a 3D radiative MHD simulation performed with the \bifrost\ code \citep{Gudiksen:2011}. This is the same simulation as was used in \citetalias{Leenaarts:Mg1} and \citetalias{Leenaarts:Mg2}. To cover a larger sample of atmospheric conditions, we used 37 simulation snapshots, including the snapshot used in \citetalias{Leenaarts:Mg2} (henceforth referred to as snapshot 385, its number in the time series).

\bifrost\ solves the resistive MHD equations on a staggered Cartesian grid. The setup used is the so-called ``box in a star'', where a small region of the solar atmosphere is simulated, neglecting the radial curvature of the Sun. The simulation we use here includes a detailed radiative transfer treatment including coherent scattering \citep{Hayek:2010}, a recipe for non-LTE (NLTE) radiative losses from the upper chromosphere to the corona \citep{Carlsson:2012}, thermal conduction along magnetic field lines \citep{Gudiksen:2011}, and allows for non-equilibrium ionization of hydrogen in the equation of state \citep{Leenaarts:2007}.

The simulation covers a physical extent of $24\times 24\times16.8$ Mm$^{3}$, extending from 2.4~Mm below the average $\tau_{500}=1$ height and up to 14.4~Mm, covering the upper convection zone, photosphere, chromosphere, and lower corona. The grid size is $504\times504\times496$ cells, with a constant horizontal resolution of 48~km and non-uniform vertical spacing (about 19~km resolution between $z=-1$ and $z=5$~Mm, up to a maximum of 98~km at the top). The photospheric mean unsigned magnetic field strength of the simulation is about 5~mT (50~G), concentrated in two clusters of opposite polarity, placed diagonally 8~Mm apart in the horizontal plane. We make use of a time series covering 30~min of solar time.  The physical quantities in the simulation were saved in snapshots every 10~s, but to keep the computational costs manageable we used only every 5$^{\mathrm{th}}$ snapshot (37 snapshots in total).

\subsection{Synthetic Spectra}

To calculate the synthetic spectra we used a modified version of the \RH\ code \citep{Uitenbroek:2001}.
 \RH\ performs NLTE radiative transfer calculations allowing for
 PRD effects, which are important in the
 \MgIIhk\ lines. As discussed in \citetalias{Leenaarts:Mg1}, full 3D
 PRD calculations with chromospheric models are not feasible at the
 moment, due to numerical instabilities (and in any case are much more
 computationally demanding). Therefore, our modified version of
 \RH\ operates under the ``1.5D'' approximation: each atmospheric column
 is treated independently as a plane-parallel 1D atmosphere. As shown in 
 \citetalias{Leenaarts:Mg1}, this is a good approximation for the
 \MgIIhk\ spectra, except at the very line core where the scattered
 radiation from oblique rays is important. Treating each column independently presents a highly parallel problem; our code is MPI-parallel and scales well to thousands of processes. Other modifications from \RH\ include the hybrid angle-dependent PRD recipe of \cite{Leenaarts:2012} and the gradual introduction of the PRD effects to ease convergence. 

For the NLTE calculations we employ the 10-level plus continuum \MgII\ atom described in \citetalias{Leenaarts:Mg1}. This atom is a larger version of the ``quintessential'' 4-level plus continuum atom used in \citetalias{Leenaarts:Mg2}. As shown in \citetalias{Leenaarts:Mg1}, the differences between both atoms are negligible for the \MgIIhk\ spectra. The larger atom was used here to allow the synthesis not only of \MgIIhk\ but also of several other weaker \MgII\ lines contained in the NUV region. 

In addition to the \MgII\ lines treated in NLTE, we also include approximately 600 lines of other species present in the NUV window of IRIS. These lines were treated assuming LTE with atomic data taken from the line lists of \cite{Kurucz:1995}. To save computational time, only the 15\% strongest lines in the $278.256-283.498$~nm\footnote{\emph{Vacuum} wavelengths are exclusively used in the text and figures.} region were used. %
The larger \MgII\ atom used and the much larger number of wavelength points necessary to cover the extra lines add to the computational expense of the problem. To keep the computational costs reasonable, we performed our calculations for every other spatial point in the horizontal directions, resulting in a horizontal grid size of $252\times 252$ or a resolution of 95~km. This spatial resolution is still more than twice as good as that of IRIS. 

\subsection{Spatial and Spectral Smearing}

In the NUV window the IRIS spectrograph has a spatial resolution of about $0\farcs4$ and a spectral resolution of about 6~pm \citep{IRIS-paper}.
To account for this finite resolution, the synthetic spectra were spatially and spectrally smeared to mimic the instrumental profile. Following the path of light, the spectrograms were first spatially smeared and then spectrally smeared. The spatial convolution was made with a preliminary point spread function (PSF) based on pre-launch calculations (De Pontieu, private communication) and similar in shape to an Airy core with $0\farcs33$ FWHM. For the spectral convolution we assumed a Gaussian profile with a FWHM of 6~pm and a spectral pixel size of 2.546~pm (corresponding to a velocity dispersion of about 2.7~$\kms$/pixel). To mimic the observational conditions the 3D spectrograms were binned into ``synthetic slits'' with a width of $0\farcs33$ and a spatial pixel size of $0\farcs166$; with our box size of $24\times 24$ Mm$^{2}$ this amounted to 100 slits of 200 spatial pixels per snapshot. 

\subsection{Simulated Slit-jaw Images}

In addition to the NUV spectrograms, IRIS will capture NUV slit-jaw images via a \v Solc birefringent filter positioned at either 279.60~nm (near the core of \MgIIk{}) or at 283.10~nm (in the far red wing of \MgIIh{}). We synthesized slit-jaw images by applying the same spatial convolution as for the spectrograms and by multiplying the 3D spectrograms by the filter transmission profiles (normalized to unit area). The filter transmission profiles are approximately Gaussian and have a FWHM of 0.38~nm. The slit-jaw images were further mapped to $0\farcs166^{2}$ pixels.

\begin{deluxetable}{lcc}
\tablecaption{Parameters of the Slit-jaw Filter Transmission Profiles\label{tab:slitjaw}}
\tablehead{
\colhead{Filter Name} & \colhead{Central Wavelength} & \colhead{FWHM} \\
     & \colhead{(nm)} & \colhead{(nm)}} 
\startdata
\MgIIk\ core    & $279.60$ & 0.38  \\
\MgIIh\ wing    & $283.10$ & 0.38  \\
BFI \CaH\       & $396.97$ & 0.22  %
\enddata
\end{deluxetable}
For comparison with the \MgII\ slit-jaw images, we also simulated slit-jaw images for \CaH\ 
to compare with the filter in the \emph{Hinode} Solar Optical Telescope (SOT) Broadband Filter Imager \citep[BFI;][]{Suematsu:2008}. \CaH\ filtergrams, in particular as observed with SOT/BFI, have been a widely used chromospheric diagnostic. To obtain the simulated \CaH\ slit-jaw images, we used the same radiative transfer code to calculate synthetic spectra using a 5-level plus continuum \CaII\ atom, allowing for PRD in the \CaH\ line. As in the \MgII\ spectra, we included the strongest atomic lines in the region, again using the line lists of \citet{Kurucz:1995}.
To simulate the BFI's \CaH\ 
filtergrams, we applied its filter transmission profiles (normalized to unit area) on the spectra. The filter transmission profiles have shapes similar to a Gaussian. The central wavelengths and approximate FWHMs for the IRIS NUV and \CaH\ BFI filters are listed in Table~\ref{tab:slitjaw}. The BFI filtergrams were convolved with the blue channel PSF of \citet{Wedemeyer2008}, scaled for the wavelength of 397~nm and mapped into $0\farcs054^2$ pixels.

\subsection{Noise\label{sec:noise}}

To estimate the effect of noise on the derived spectral quantities, we added artificial noise to the synthetic spectra. For simplicity and because the final noise profile of the IRIS observations is not yet known, we have assumed that photon noise dominates and thus consider pure Poisson noise. After the spatial and spectral convolution, different amounts of noise were added to obtain spectra with different mean signal-to-noise ratios (S/Ns).

Given the intensity variations over the NUV window, different spectral lines will have different S/N levels. For this work we defined the mean S/N level as the S/N at a wavelength of 280.042~nm, approximately the maximum intensity point between the h and k lines. According to the spatially and temporally averaged spectrum, wavelengths below 279.31~nm and above 280.59~nm will have a larger S/N. At the highest point of the k line the S/N in the synthetic spectra is on average about 44\% higher, and about 23\% higher for the h line. Near the end of the NUV window at 283~nm the S/N is about three times higher than at 280.042~nm.

To achieve a given S/N, the spatially and temporally averaged spectrum was scaled so that the intensity at 280.042~nm was equal to the square of the target S/N level. The intensity at each point was randomly taken from a Poisson distribution with an expected value equal to the scaled intensity. Finally, the intensities were scaled back to the initial units. To cover a wide range of observing possibilities, six levels of S/N were used: 100, 50, 20, 10, 5, and 2. 

\begin{figure}
\begin{center}
\includegraphics[scale=0.87]{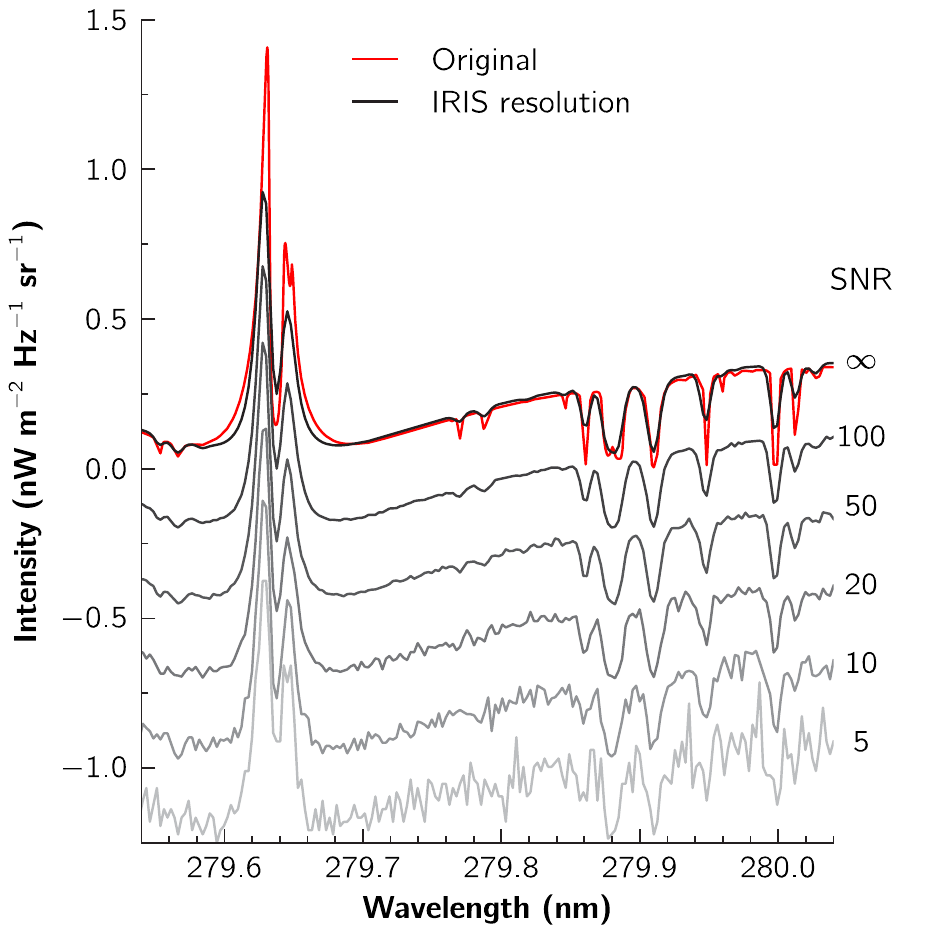}
\end{center}
\caption{Effects of IRIS spatial and spectral resolution and noise on the NUV spectrum near the \MgIIk\ line. The original spectrum was taken from a single simulation pixel. This pixel was brighter than its surroundings and was chosen so that the \MgIIk\ intensity was similar to the mean observed value. The S/N is measured at the last wavelength shown (280.042~nm). For clarity, spectra of different S/N are offset 0.25 nW~m$^{-2}$~Hz$^{-1}$~sr$^{-1}$ from each other. \label{fig:noise_spec}}
\end{figure}

The effects of spatial and spectral resolution and noise are illustrated in Fig.~\ref{fig:noise_spec}. Compared with the original spectrum, one can see how the instrumental resolution of IRIS affects the widths of lines and how in this example the spatial resolution lowers the \MgIIk\ peak intensities and the overall intensity level. The effects of noise on \MgIIk\ start to be noticeable at a S/N of 20.

\section{Mg II h \& k diagnostics}  \label{sec:mg}

\subsection{Extracting Spectral Properties\label{sec:mg_extract}}

\begin{figure}
\begin{center}
\includegraphics[scale=0.87]{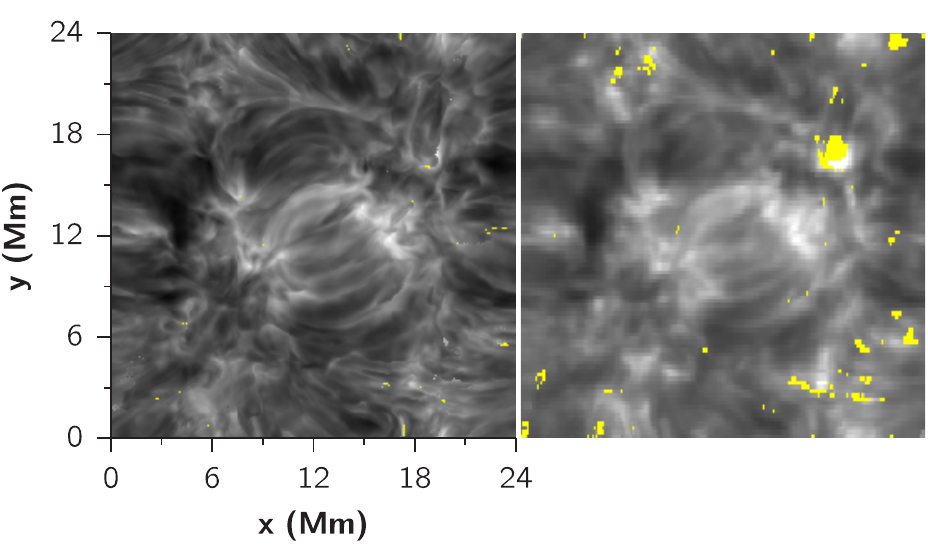} 
\end{center}
\caption{Effects of IRIS spatial and spectral resolution on the \kthree\ intensity for snapshot 385. \emph{Left:} \kthree\ radiation temperature extracted from original spectra. \emph{Right:} \kthree\ radiation temperature extracted from spectra convolved with the instrumental profile. Brightness is linearly scaled from 3.9 to 5.3~kK. Yellow points are locations where the detection algorithm failed to find \kthree. \label{fig:k3int}}
\end{figure}

The solar spectra of each of the \MgIIhk\ lines are often described by a central absorption
core surrounded by two emission peaks, in turn surrounded by local
minima. These features are called  \hthree{}, \htwoV/\htwoR{}, and  \honeV/\honeR{} respectively for the h line, and similarly for the k line. The locations of the spectral features were extracted from the spectra using an automated procedure. Given the sheer number of spectra analyzed (about 11 million), a manual feature identification was not possible. Using a peak finding procedure, we developed an algorithm to automatically extract the locations of the \hthree/\kthree\ line cores and the \htwo/\ktwo\ maxima. The algorithm identifies how many peaks exist in the spectrum and determines, according to a set of rules (detailed in \citetalias{Leenaarts:Mg2}), which features are the emission peaks and the central absorption core.

\begin{figure*}
\begin{center} 
\includegraphics[scale=0.85]{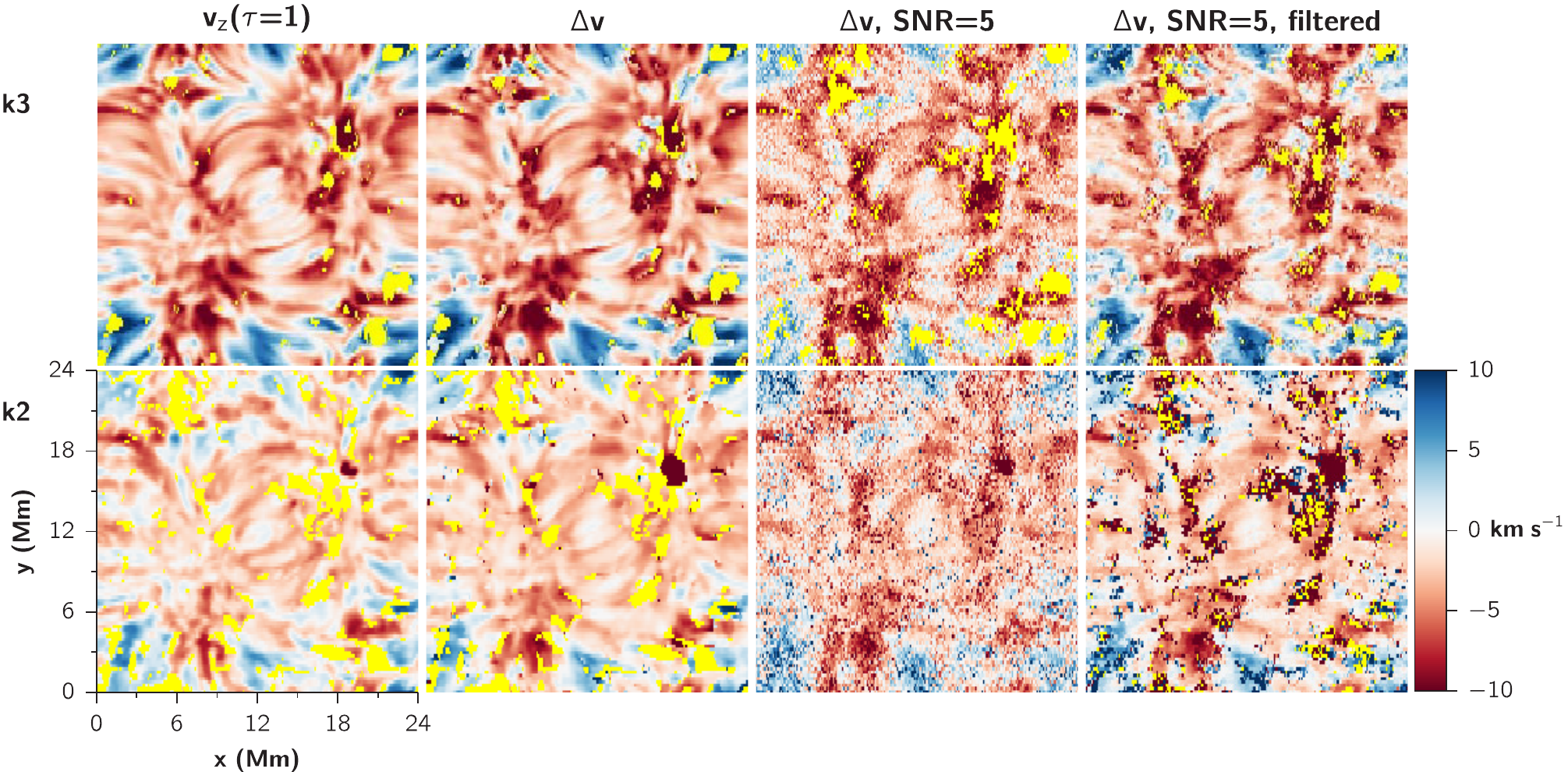}
\end{center}
\caption{Velocity maps for the \kthree\ line core (\emph{top panels}) and \ktwo\ peaks (\emph{bottom panels}). The color scale is clipped at $-10\;\kms$ and $10\;\kms$ to maximize contrast. Yellow points represent locations where the detection algorithm failed to find the features. The first column shows the atmospheric velocity at the feature's $\tau=1$ depth. The remaining columns show the velocity shift of the feature as detected from the spectra, for spectra without noise, with noise at S/N of 5, and with noise at S/N of 5 but using a Wiener filter (see the text), respectively.\label{fig:mg_vz}}
\end{figure*}

The parameters for the feature detection algorithm were adjusted to work
better with data at the resolution of IRIS. The
convolved spectra were interpolated to a higher
resolution of $\approx0.3$~pm, from the 2.546~pm pixels of IRIS. As discussed
in \citetalias{Leenaarts:Mg2}, the \MgIIhk\ synthetic profiles from
this simulation are not as strong or wide as observed (see
Figure~\ref{fig:mspec_lines}). At the spectral resolution of IRIS, a
larger proportion of synthetic spectra will have the
\htwo/\ktwo\ peaks blended as one feature, with no visible line core depression. To compensate for this, the detection algorithm was improved to work with these cases (about 6\% of our spectra). When no local minimum is visible between the blue and red peaks of the k or h lines, the location of the line core is taken as the point where the $\left|\mathrm{d}I/\mathrm{d}\lambda\right|$ is lowest in the interpeak region. Because in these cases the location of one of the peaks is not known, $\left|\mathrm{d}I/\mathrm{d}\lambda\right|$ is evaluated in an interval with a length of $\approx10\:\kms$ (roughly the lowest measurable peak separation), starting about 4.6$\;\kms$ blueward of the red peak or redward of the blue peak. This reduced interval was chosen to avoid the derivative from being evaluated over the red and blue peaks, were the zero derivative would cause a spurious line core detection. This extra step of taking the minimum of $\left|\mathrm{d}I/\mathrm{d}\lambda\right|$ is done only for the spectra with blended peaks; visual inspection showed that it works as intended in more than 70\% of the cases with blended peaks.

After extracting the positions of the spectral features, our goal was to correlate them with physical quantities in the atmosphere. To find the corresponding atmospheric quantities, the first step was to upscale (using nearest-neighbor interpolation) the wavelengths of the spectral features from the synthetic spectrogram grid to the simulation grid. This was necessary because the synthetic spectra have the IRIS spatial pixels of $0\farcs166$ with a slit width of $0\farcs33$, and the atmospheric quantities are stored in the simulation's grid, which has a higher spatial resolution. Afterwards, the atmospheric properties were extracted from each column in the simulation. The optical depths from the simulation were interpolated to the wavelengths of the observed features and, for each spectral feature $z(\tau=1)$, the height where the optical depth reaches unity was calculated. Simulation variables such as vertical velocity and temperature were then extracted for each column at the height given by $z(\tau=1)$. In some cases, we calculated statistics on atmospheric properties between two $z(\tau=1)$ heights from different spectral features. For a given spectral feature, we extracted one value of each atmospheric quantity for every column in the simulation. This resulted in 2D maps $(x,y)$ for these quantities. For a meaningful comparison with the spectral properties, the resulting 2D maps of each atmospheric quantity were then spatially convolved using the same procedure as for the synthetic slits.

In Figure~\ref{fig:k3int} we show the effects of the IRIS spatial and
spectral resolution on the measured \kthree\ intensity, for
snapshot 385 (calculated in full 3D CRD, from \citetalias{Leenaarts:Mg2}). As
expected, the very fine structure is washed out, but in general the
large-scale structure closely reflects what is seen in the original
spectra. The correlation between \kthree\ intensity and $z(\tau=1)$
heights is, however, largely lost. Even at full spectral and spatial
resolution, the Pearson correlation coefficient was only $-0.39$. The
finite spatial and spectral resolution of IRIS leads to a difference in
the measured intensity and Doppler shift of \kthree\ and their true
values at the resolution of the simulation. Owing to the narrow
extinction profile (the Doppler width in the simulation is typically
2.5~km~s$^{-1}$), small differences between measured and true Doppler
shifts lead to large variations in the associated $z(\tau=1)$
height. Based on this simulation snapshot, we conclude that the
\kthree\ intensity cannot be blindly used as measure of the variation
of the $z(\tau=1)$ height.

Nevertheless, some properties of the correlation are retained at IRIS
resolution: the very brightest pixels in the IRIS resolution image are
still formed deepest in the atmosphere, and pockets of chromospheric
material that extend high up into the corona still have a low
intensity. We also note that the simulated profiles have a narrower
central depression than observed. Therefore it might be that real
observations are less sensitive to the uncertainty in the Doppler shift
of the \kthree\ minimum than the simulation, and thus retain a better
intensity- $z(\tau=1)$ height correlation at IRIS resolution.

\subsection{Velocities}

Given their peculiar profile, the \hk\ spectra are rich in velocity diagnostics. Here we focus on the most important: the velocity shift of \kthree, the \ktwo\ mean velocity shift, the k peak separation, and the ratio of \ktwo\ intensities. The corresponding quantities for the h line behave similarly and were not included for brevity. As shown in \citetalias{Leenaarts:Mg2} these spectral features are related to different properties of the atmosphere.

\begin{figure*}
\begin{center} 
\includegraphics[scale=0.87]{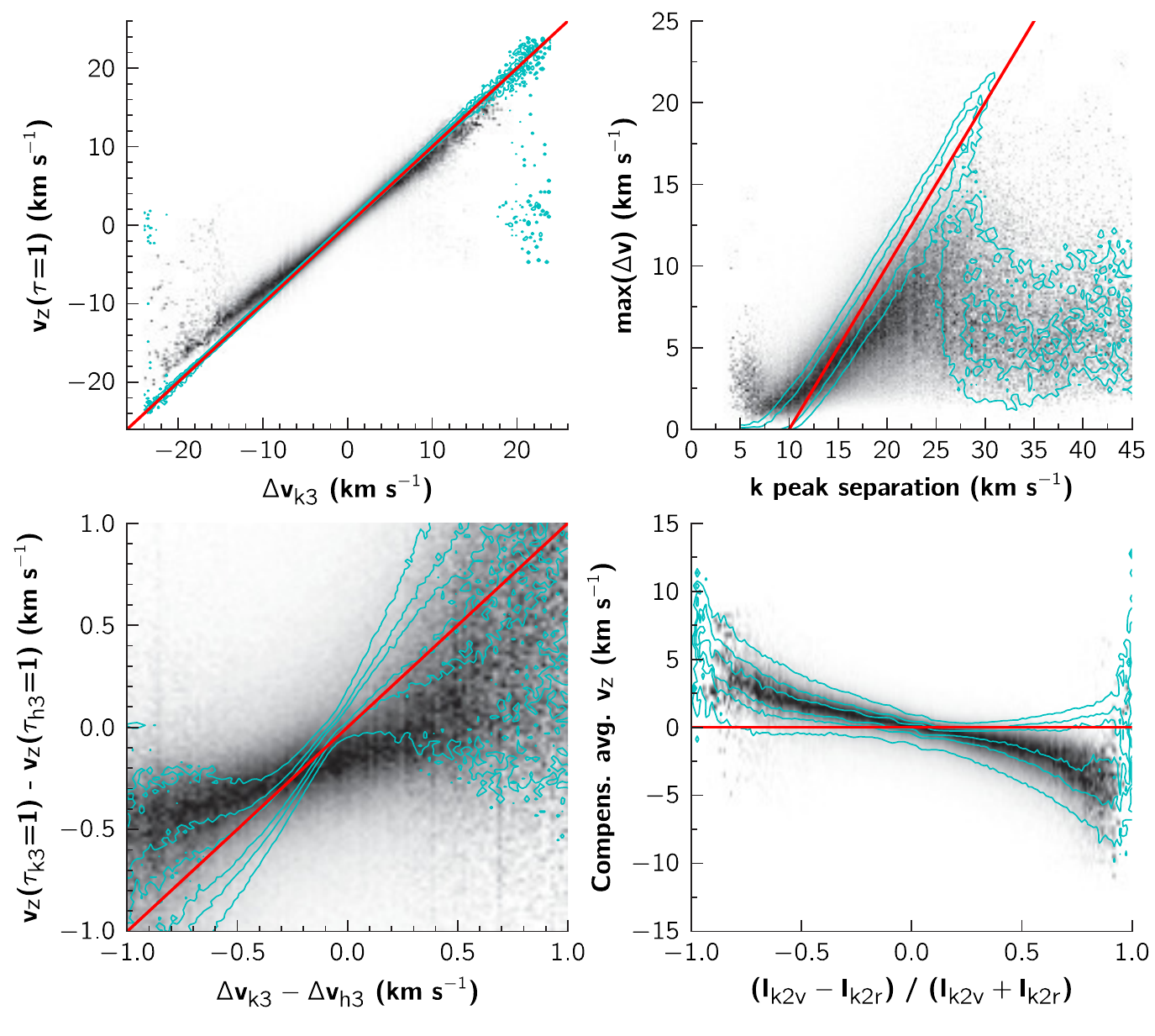}
\end{center}
\caption{Probability density functions for different atmospheric properties (y axes) for different values of the observed quantities (x axes), comparing the data extracted from spectra convolved with the instrumental profile of IRIS (\emph{grayscale}, darker means more frequent), and for the original spectra (\emph{cyan contours}). \emph{Top left panel: } \kthree\ velocity shift vs. atmospheric velocity at the corresponding height where $\tau$=1; red line denotes $y=x$. \emph{Bottom left panel:} difference between \kthree\ and \hthree\ velocity shifts vs. the difference in corresponding $\tau$=1 atmospheric velocities; red line denotes $y=x$. \emph{Top right panel:} k peak separation vs. the maximum atmospheric velocity extremes between the formation height of the \ktwo\ and \kthree{}; red line denotes $y=x$$-$10. \emph{Bottom right panel:} intensity ratio of \ktwo\ peaks vs. the compensated average $v_z$ (see text); red line denotes $y=0$.\label{fig:mg_vel}}
\end{figure*}

In Figure~\ref{fig:mg_vz} we show a comparison of velocity maps for snapshot 385. Throughout this paper we adopt the convention that positive velocities correspond to upflows (or blueshifts). The atmospheric velocities $v_z(\tau=1)$ were taken from the simulation at the heights where each feature reaches an optical depth of unity (i.e., a measure of velocity at the typical height of formation) and are compared with $\Delta v$, the velocity shift of the spectral features (the observable). In the case of \ktwo{}, both $v_z(\tau=1)$ and $\Delta v$ are the average of the red and blue peaks, only when both peaks were detected. Cases where the automated feature detection could not find the spectral features are shown in yellow. These are more frequent for \ktwo\ because both peaks had to be detected to measure the average velocity, and there are many spectra where only one peak is seen. The second column of Figure~\ref{fig:mg_vz} shows $\Delta v$ from the spectra convolved to the IRIS resolution, but with no added noise. The third column shows the resulting $\Delta v$ when including a substantial amount of noise (mean S/N of 5), and the fourth column shows the results when using a noise filter on the spectra before extracting $\Delta v$ (noise filtering is discussed in Section~\ref{sec:mg_noise}).

From Figure~\ref{fig:mg_vz} one can see that there is a very good correlation between the atmospheric $v_z(\tau=1)$ and the $\Delta v$ velocity shifts for both \kthree\ and \ktwo. The overall morphology and sign still hold even with a substantial amount of noise, although some fine structure is lost. The mean $\tau=1$ height for \kthree\ is about 2.5~Mm, meaning its velocity shifts are closely related to the velocity in the upper chromosphere. The mean $\tau=1$ height for the \ktwo\ peaks is about 1.5~Mm, meaning that their $\Delta v$ brings complementary velocity information at a lower chromospheric height. The vertical velocity increases with height in the simulated chromosphere and therefore the \ktwo\ velocity shifts are lower than those of \kthree. The \ktwo\ velocities also appear more vulnerable to noise than those of \kthree.

In Figure~\ref{fig:mg_vel} we show a more quantitative view of the different velocity diagnostics, showing also the effect of the instrumental resolution. In these figures, we plot the observed quantities on the x axes. To better display the relations we build 2D histograms from the scatter plots and scale them by the maximum value for each value on the x axis, creating a probability density function (PDF) for a given atmospheric quantity when the observed quantity is given by the value on the x axis. This illustrates the quantity distribution for the range of observable values in the x axis and improves the visibility of the relation for areas were the point density is lower. Using data for all the 37 snapshots, each of these PDFs comprises more than $10^6$ points. The distributions for the quantities extracted from the IRIS-convolved spectra are shown in greyscale (darker corresponds to higher probability); the distributions for the original spectra are shown in cyan contours. The contour levels were chosen so that the lowest contour seen has approximately the same level as the lightest grey visible in the greyscale images.

The upper-left panel of Figure~\ref{fig:mg_vel} shows the relation between $\Delta v_{\mathrm{k3}}$ and its $v_z(\tau=1)$, which confirms the very good correlation seen in Figure~\ref{fig:mg_vz}. For the original spectra the agreement is excellent, with a very tight correlation with little scatter. This very good agreement still holds when taking into account the resolution of IRIS, but two effects are seen: the range of velocities is not as large and there is a slightly changed tilt in the distribution: higher shifts tend to correspond to lower atmospheric velocities. The smoothing of the extreme values is mostly a consequence of the limited spatial resolution. The changed tilt of the distribution, on the other hand, is mostly due to the limited spectral resolution and pixel size. In absolute terms, the highest shifts in velocity tend to occur in points where one of the \ktwo\ peaks is much stronger than the other or where there are several close local minima in between the peaks. In such cases the spectral convolution smooths the line profiles in a way that causes the \kthree\ minima to become farther away from the original value or a smoothed average of the several local minima. This causes the observed shift to be larger, which does not correspond as well to the atmospheric velocities and causes the changed tilt of the distribution. In any case, the effects of the instrument resolution on $\Delta v_{\mathrm{k3}}$ are very small (hardly noticeable in Figure~\ref{fig:mg_vz}).

The lower-left panel of Figure~\ref{fig:mg_vel} shows the relation between the difference of the \kthree\ and \hthree\ velocity shifts against the difference in their corresponding $v_z(\tau=1)$. The difference in $z(\tau=1)$ between \kthree\ and \hthree\ is a few tens of km \citepalias{Leenaarts:Mg2}. Therefore, the difference in their velocity shifts is proportional to the difference in atmospheric velocities between their formation heights. This relation has a substantial amount of scatter for $\left|\Delta v_{\mathrm{k3}}-\Delta v_{\mathrm{k3}}\right| \gtrsim 0.5\:\kms$. This is mostly because of reduced statistics: only $\approx 12$\% of the points have an absolute velocity difference larger than 0.5~$\kms$ (and about 5\% have a velocity difference larger than 1~$\kms$). In any case, the results show that at the resolution of IRIS this relation can be used to detect the sign of the velocity difference and obtain an estimate of the vertical acceleration of chromospheric material. With spectra at the resolution of IRIS one gets lower velocity differences and a scatter that is about three times larger than that of the original data. The slight change in tilt observed in the distribution of $\Delta v_{\mathrm{k3}}$ versus $v_z(\tau=1)$ is more pronounced here because it is derived from the \kthree\ and the \hthree\ quantities, both of which suffer from the effects discussed in the previous paragraph.

The upper-right panel of Figure~\ref{fig:mg_vel} shows the relation between the k peak separation and $\max(\Delta v)$, the absolute value of the difference between the maximum and minimum atmospheric velocities in the formation region of the interpeak profile (the range between the mean $\tau=1$ height of \ktwoV\ and \ktwor\ and the $\tau=1$ height of \kthree). This is an important diagnostic that relates an easy-to-measure quantity (the peak separation) with the large-scale velocity gradients in the upper chromosphere. The correlation is better seen in the original data; the peak separation is the velocity diagnostic most affected by the spatial and spectral resolution. As seen in Figure~\ref{fig:mg_vel}, the effects of the IRIS resolution are threefold: more extreme values of peak separation are no longer seen, there is a larger scatter in the distributions, and there is a slight change in the tilt of the previously near-linear correlation with $\max(\Delta v)$. For peak separations larger than 25~$\kms$ the correlation is lost, even in the original data. The main reason is because the velocities in the simulation are not as violent as the Sun -- one would expect the trend to continue if larger peak separations were common in the synthetic spectra. From the original spectra, only about 6\% of the points have peak separations larger than 25~$\kms$ (5\% in the spectrally convolved spectra). The origin of these points, discussed in \citetalias{Leenaarts:Mg2}, comes from either misidentifications in the detection algorithm or points where the broadening is not due to velocity effects but from temperature maxima taking place deeper in the chromosphere, causing the \ktwo\ and \htwo\ peaks to be formed deeper and their $\tau=1$ being reached further from the line core. The same effect is visible in the IRIS-convolved spectra.

The difference in tilt of the peak separation distribution from the linear relation with $\max(\Delta v)$ and its increased scatter are a consequence of the instrumental resolution. This deviation happens mostly because the extinction profiles fall sharply outside the line cores (see Figures 9--11 of \citetalias{Leenaarts:Mg2}). A small error in the velocity shift of the peaks can result in its estimated $z(\tau=1)$ being wrong by several Mm. This in turn causes the $\max(\Delta v)$ to be evaluated at different atmospheric layers than those that contributed to the peak separation, leading to a greater scatter in the relation. The effect is amplified because two measurements are needed for the peak separation.

In our forward modelling approach we did not add
  microturbulence to the computation of the synthetic profiles. The
  peak separation is thus solely set by the simulation properties, and is
  smaller than the observed separation (see Figure~\ref{fig:mspec_lines}). Because of the reduced peak separation, the effects of IRIS spectral resolution in the synthetic peak separation reflect a worst-case scenario. \citet{Staath:1995} report a k peak separation of 31~pm ($\approx 33\;\kms$), which is much larger than in most of our spectra. We expect the effects of the IRIS spectral resolution to be less severe for real data.

The lower-right panel of Figure~\ref{fig:mg_vel} shows the relation between the intensity ratio of the blue and red peaks,
\begin{equation}
R_\mathrm{k}\equiv\frac{I_\mathrm{k2v}-I_\mathrm{k2r}}{I_\mathrm{k2v}+I_\mathrm{k2r}},
\end{equation}
and the compensated average $v_z$, defined as the mean velocity
between the heights of formation of \ktwo\ and \kthree{} minus the
mean velocity at the \ktwo\ height of formation. As discussed in
\citetalias{Leenaarts:Mg2}, this intensity ratio provides an important
diagnostic of atmospheric velocity above the peak formation height: a
stronger blue peak means that material above the \ktwo\ height of
formation is moving down and a stronger
red peak means material is moving up. As shown in
Figure~\ref{fig:mg_vel}, this diagnostic remains robust when taking into
account the instrumental smearing. As expected, the spatial resolution
of IRIS causes the more extreme values to be smoothed out, but the
overall relation is unchanged. For this diagnostic, the distribution
of the IRIS-convolved data shows in fact less scatter than that of the
original data. In some cases this happens because the spectral
convolution smooths more complex peak structures (some with multiple
sub-peaks in \ktwoV\ or \ktwoR) into an average value, leading to less
scatter in $R_\mathrm{k}$. A limitation of $R_\mathrm{k}$ is
  that it is only defined for spectra with two peaks. 
 In the case of single-peaked profiles, one could instead take intensities over a wavelength window around the peaks \citep[e.g., see][]{Rezaei:2007}.

\subsection{Temperatures}

\begin{figure}
\begin{center} 
\includegraphics[scale=0.84]{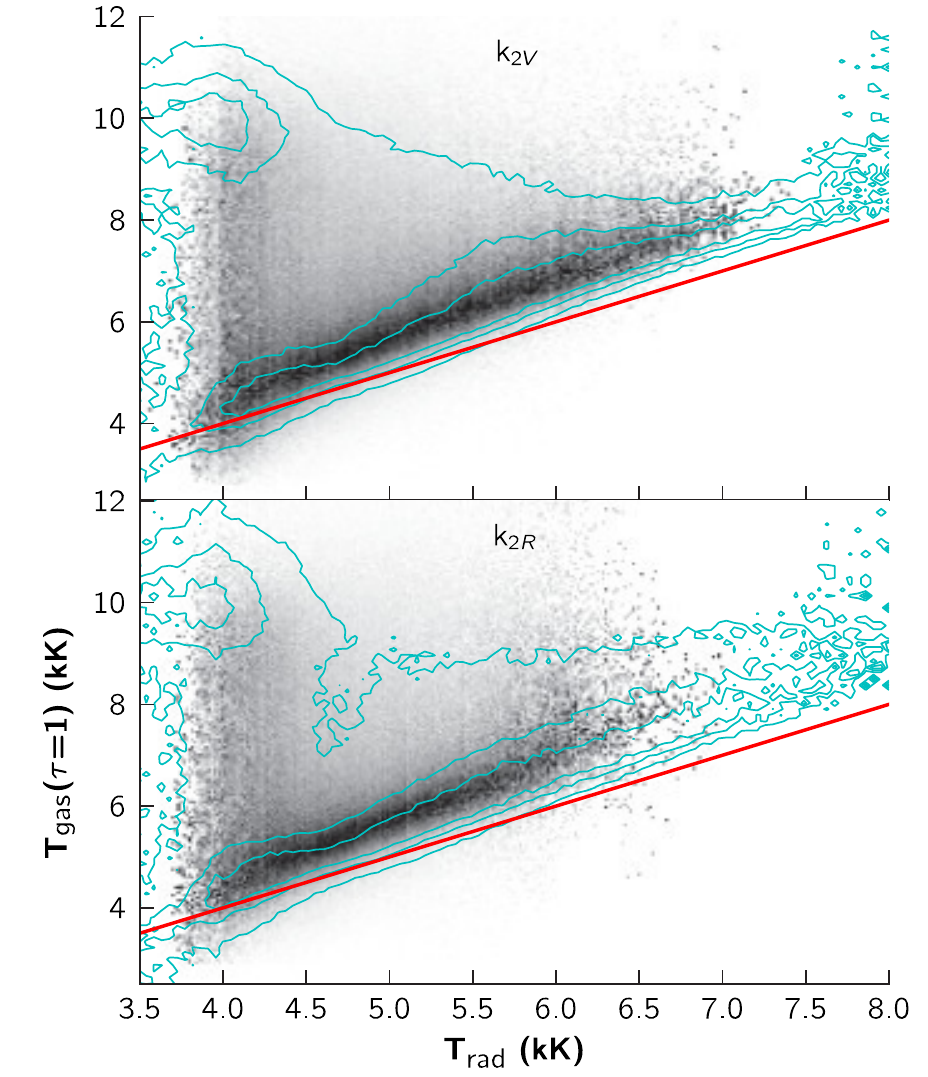}
\end{center}
\caption{PDFs for the gas temperature at optical depth unity being $T_{\mathrm{gas}}$ when the observed \ktwoV\ (\emph{top}) and \ktwoR\ (\emph{bottom}) radiation temperatures are $T_{\mathrm{rad}}$, extracted from spectra convolved with the instrumental profile of IRIS (\emph{grayscale}, darker means more frequent), and for the original spectra (\emph{cyan contours}). The red line denotes $y=x$.\label{fig:mg_temp}}
\end{figure}

Besides the velocity information, the \MgIIhk\ lines also provide
estimates of atmospheric temperatures. At the \kthree\ and
\hthree\ cores the source function is decoupled from the local
conditions, but some coupling remains at \ktwo\ and \htwo, which are
formed deeper, in the middle chromosphere
\citepalias{Leenaarts:Mg2}. The intensity from the \ktwo\ and
  \htwo\ peaks can be converted to $T_\mathrm{rad}$, the radiation (or
  brightness) temperature, which can thus be used as a proxy for
$T_\mathrm{gas}(\tau=1)$, the gas temperature at their heights of
formation. The conversion of the synthetic spectra to
  $T_\mathrm{rad}$ can be done directly because the computation yields
  absolute intensities, while
  for IRIS data one needs to perform an absolute calibration of the
  spectra first.

In Figure~\ref{fig:mg_temp} we show the PDFs for $T_\mathrm{gas}(\tau=1)$ given an observed $T_\mathrm{rad}$ of \ktwoV\ and \ktwoR. Again, these PDFs have been scaled by the maximum in each column along the x axes. There is some scatter in this relation, but nevertheless it is reasonable for a large number of points. The radiation temperature of the peaks is a strong constraint on the minimum atmospheric temperature at the peak formation height (middle chromosphere), given that $T_\mathrm{gas}(\tau=1)$ is seldom less than $T_\mathrm{rad}$.

Our results indicate that the effect of instrumental smearing does not influence the correlation significantly. The finite spatial and spectral resolution smooths out the more extreme intensity values (in particular, above 7~kK), but the correlation is mostly unchanged. The \ktwoR\ peak seems more affected by the narrower range of $T_\mathrm{rad}$. In the original data (as noted in \citetalias{Leenaarts:Mg2}), there is a cluster of points at low $T_\mathrm{rad}$ for which $T_\mathrm{gas}(\tau=1)$ is about 10~kK. In \citetalias{Leenaarts:Mg2}, these are identified as points where the radiation comes mostly from upper layers in the chromosphere, where the source function has decoupled from the local temperature. With the instrumental smearing, the low $T_\mathrm{rad}$ values of such points are blurred, and this localized cluster of high $T_\mathrm{gas}(\tau=1)$ disappears.

\subsection{Noise Mitigation\label{sec:mg_noise}}

The \MgII\ spectral features are related to properties of the atmosphere, but how do the relations hold when one considers the effects of noise? With many observing modes and a focus on fast rasters, IRIS will provide data with varying levels of noise. In our investigation, we want to identify the minimum S/N that can be used to derive meaningful relations from \MgIIhk\ spectra and to identify ways to mitigate the effects of noise.

We ran the automated detection algorithm on spectra with Poisson noise added, as detailed in Section~\ref{sec:noise}. To mitigate the effects of noise, we also studied the effects of applying a Wiener filter \citep{Wiener:1949} on the noisy spectra. A Wiener filter is the optimal filter for Gaussian noise and signal and seems to work well on our synthetic spectra with Poisson noise. To mimic the data products from IRIS, the Wiener filter was applied on each 2D ($y$, $\lambda$) spectrogram slice that comprises our 3D ($x$, $y$, $\lambda$) spectra (in the right panels of Figure~\ref{fig:mg_vz} this can be seen by noting how the noise patterns seem to be aligned in the horizontal direction).
The specific details of the filter are not important; its function here is to show that more can be extracted from noisy spectra -- one could also devise other Fourier filters to work on noisy data (see, e.g., the low-pass filter used by \citealt{Pereira2009b}).

\begin{figure}
\begin{center} 
\includegraphics[scale=0.84]{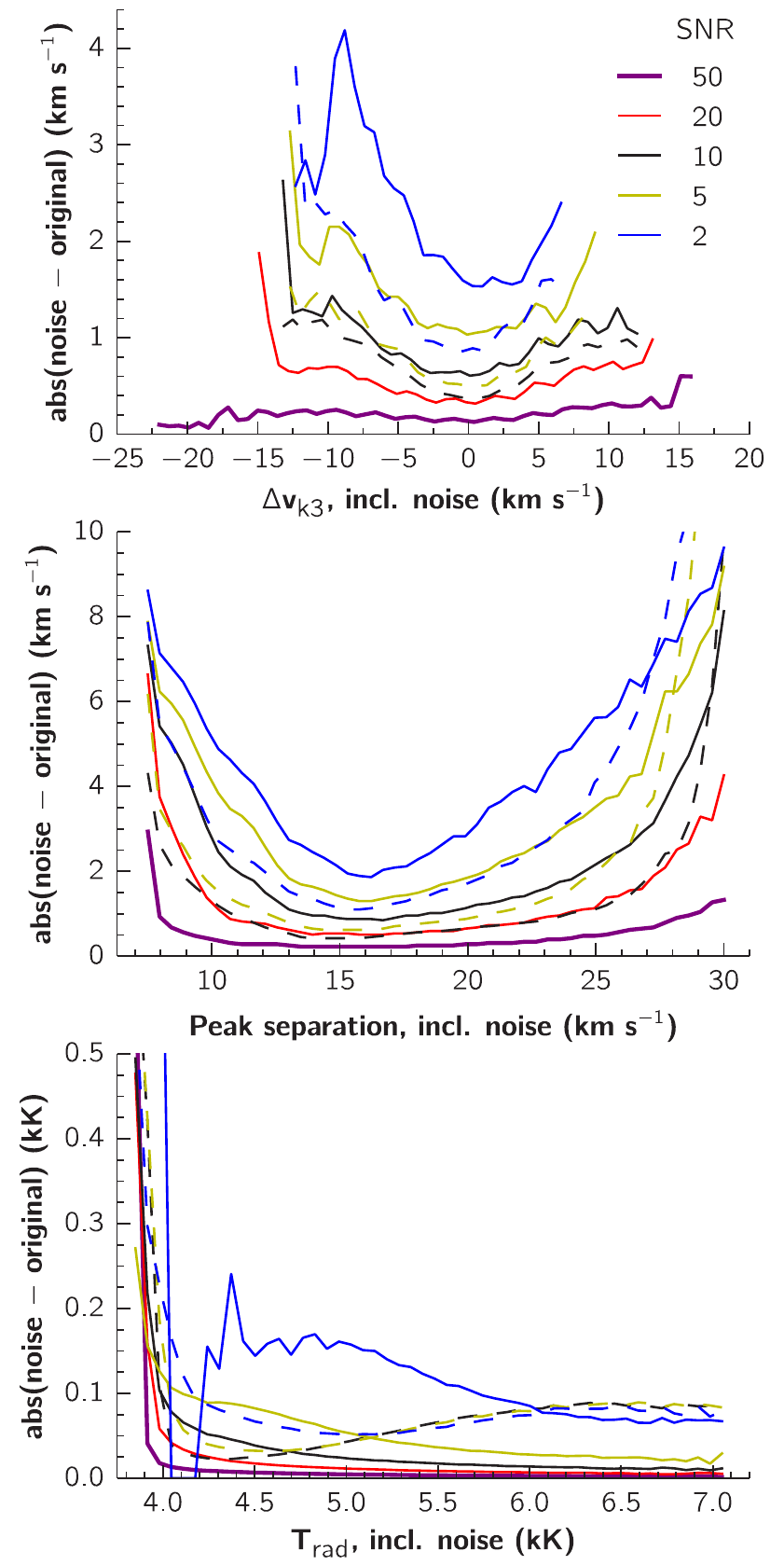}
\end{center}
\caption{Median values of the absolute difference between quantities derived from spectra with and without noise, for bins in the quantities in the x axes. We show results for different S/Ns, as per the legend. The dashed lines of the same color (only shown for S/Ns of 2, 5, and 10) show the results for the noisy spectra when a Wiener filter was applied (see text). \emph{Top panel:} effect of noise on the \kthree\ velocity shifts; \emph{middle panel:} effect of noise on the \ktwo\ peak separation; \emph{bottom panel:} effect of noise on the \ktwoV\ radiation temperature.\label{fig:mg_noise}}
\end{figure}

In Figure~\ref{fig:mg_noise}, we compare the differences between the quantities extracted from spectra with and without noise (all spectra were spatially and spectrally convolved with the instrumental profile). The different lines in Figure~\ref{fig:mg_noise} show, for a given value of a quantity measured from the noisy spectra, the location of the median of the absolute value of the difference between the ``noisy'' and ``original'' quantities. That is, if plotted as a scatter plot the lines show the point for a given absciss\ae\ bin where half of the points are above and half are below -- a measure of the difference between the quantities derived with and without noise. This is shown for three key diagnostics: \kthree\ velocity shift, peak separation, and \ktwoV\ intensity. For the S/N levels of 2, 5, and 10 the dashed lines show the results for the noisy spectra with the Wiener filter applied. At a S/N of 100 or higher the effects of noise are negligible and are not shown for clarity.

The top panel of Figure~\ref{fig:mg_noise} shows the noise effects on
$\Delta v_{\mathrm{k3}}$, the \kthree\ velocity shift. With increasing
noise levels, two effects are visible: an increasing median difference
from the original values, and a narrower range of velocity shifts
found. These reflect the increasing difficulty of detecting the
\kthree\ position when noise seems to shift the location of the line
core or add spurious peaks in its region. The narrower range of
$\Delta v_{\mathrm{k3}}$ happens because when faced with multiple
central line depressions, our detection algorithm tends to prefer those that represent smaller shifts. Higher median differences at the extreme $\Delta v_{\mathrm{k3}}$ mean that the correlation with $v_z(\tau=1)$ departs from linear, and starts to become flatter (i.e., weaker correlation). Nevertheless, even without filtering the $\Delta v_{\mathrm{k3}}$ diagnostics are moderately resilient to noise. The effects of noise start to become 
pronounced at a S/N of 5 (see Figure~\ref{fig:mg_vz}), and are much stronger at a S/N of 2. %

A Wiener filter on the noisy spectra ameliorates significantly the effects of noise. With the filtered data, results for a S/N of 2 improve the agreement with that of unfiltered S/N $=$ 5, and similarly from S/N~$= 5-10$. At a S/N of 10 the improvement of the filter is proportionally smaller, and at S/Ns of 20 or above (not shown) it is difficult to obtain an accurate estimate of the noise when constructing the filter and consequently it may actually be
counterproductive to apply a Wiener filter -- not only does the filter not improve the results, it actually increases 
the discrepancy because it tries to compensate for noise that is not there. For all of the diagnostics, it seems that the results at a S/N of 20 are the limit of what the Wiener filter can improve.

The middle panel of Figure~\ref{fig:mg_noise} shows the noise effects on the peak separation. The peak separation is more vulnerable to noise because it needs two spectral quantities. Above a median difference of 2~$\kms$ the diagnostic potential of the peak separation is lost. Without a noise treatment, the limiting S/N for an acceptable peak separation estimate is around 10. As seen in Figure~\ref{fig:mg_noise}, the Wiener filter seems to work very well at improving the peak separation estimates, and a S/N of 5 still provides reasonable results. 

Finally, the bottom panel of Figure~\ref{fig:mg_noise} shows the effects of noise on the \ktwoV\ intensities, converted to $T_{\mathrm{rad}}$. Of the tested quantities, these are the most resistant to noise. The lowest intensities ($T_{\mathrm{rad}} < 4$~kK) are much more affected because these points have a much weaker signal: at a given mean S/N level, their effective S/N is much lower. Outside of these regions, the effects of noise are very small -- compared to the scatter in the distribution on Figure~\ref{fig:mg_temp} they are negligible. The gap in $T_{\mathrm{rad}}$ for a S/N of 2 is a consequence of the large amount of noise: at this level the noise is so high that the intensity at these lower values becomes quantized, and this gap is a consequence of that quantization. The Wiener filter corrects this problem, but given that the noise effects are generally so weak for $T_{\mathrm{rad}}$, it is unnecessary for other S/N levels.

\section{Photospheric diagnostics}             \label{sec:phot}

\subsection{Additional Information in NUV Spectra}

Besides the \hk\ lines, the NUV is rich in other spectral lines and they provide a wealth of additional information. Here, we focus on velocities and temperatures derived from spectral regions in the NUV window.

The many lines blended in the wings of \hk\ have a great diagnostic potential, but at the same time the large number of these lines means that virtually every line has multiple blends. Most of the lines in the NUV window are formed in the photosphere, at varying heights of formation. Multi-line velocity diagnostics are a powerful tool for tracing the vertical velocity structure and have been used in many chromospheric and photospheric studies \citep[e.g.,][]{Lites:1993, Beck:2009, Felipe:2010}.

Given the additional lines we included in our line synthesis, we sought to identify which lines were in cleaner spectral regions and from which we could extract reliable velocity estimates. With the help of the observed spectra from the RASOLBA balloon experiment \citep{Staath:1995} and the quiet Sun data from the HRTS-9 rocket \citep{Morrill:2008}, we identified lines in our synthetic spectrum that matched observed features and were seen in relatively clear spectral regions. While the RASOLBA spectrum has an excellent resolution and S/N, it unfortunately only covers a small wavelength region. The spectrum of HRTS-9 covers a larger region, but with lower spectral resolution. 

\begin{deluxetable}{lrrrrc}
\tablecaption{Selected Photospheric Lines for Velocity Estimation\label{tab:lines}}
\tablehead{
\colhead{Species} & \colhead{$\lambda_0$} & \colhead{$\log gf$} &  \colhead{$E_{\mathrm{low}}$} & \colhead{$z$} & \colhead{$\sigma(\Delta v)$} \\
     & \colhead{(nm)} & & \colhead{(cm$^{-1}$)} & \colhead{(Mm)} & \colhead{($\kms$)} }
\startdata
Cr\,\textsc{ii} & 278.7295   & $-0.099$   & 38396.230 & $0.17\pm 0.02$ & 0.783 \\
C\,\textsc{i}   & 281.0584   & $0.389$    & 47957.045 & $0.17\pm 0.03$ & 0.827 \\
Ni\,\textsc{i}  & 281.5179   & $-0.033$   & 27260.891 & $0.17\pm 0.03$ & 0.838 \\
Ni\,\textsc{i}  & 281.6010   & $-2.375$   & 13521.352 & $0.17\pm 0.03$ & 0.672 \\
\hline \\[-1.5ex]
Ti\,\textsc{ii} & 278.5465   & $-1.820$   & 4897.650  & $0.22\pm 0.03$ & 0.747 \\
Cr\,\textsc{ii} & 278.6514   & 0.290      & 33618.940 & $0.21\pm 0.03$ & 0.756 \\
Fe\,\textsc{i}  & 280.9154   & $-2.431$   &  7728.059 & $0.28\pm 0.05$ & 0.581 \\
\hline \\[-1.5ex]
Fe\,\textsc{i}  & 279.2327   & $-1.769$   & 23711.454 & $0.38\pm 0.03$ & 0.386 \\
Fe\,\textsc{i}  & 280.6634   & $-2.652$   & 11976.238 & $0.38\pm 0.03$ & 0.749 \\
\hline \\[-1.5ex]
Fe\,\textsc{i}  & 280.6897   & $-1.789$   & 18378.185 & $0.42\pm 0.03$ & 0.608 \\
\hline \\[-1.5ex]
Fe\,\textsc{i}  & 279.3223   & $-1.910$   & 12560.933 & $0.50\pm 0.04$ & 0.659 \\
Cr\,\textsc{ii} & 280.1584   & 0.560      & 33694.150 & $0.58\pm 0.03$ & 0.842 \\
Fe\,\textsc{i}  & 280.5690   & $-0.703$   & 21999.129 & $0.58\pm 0.03$ & 1.093 \\
Ni\,\textsc{i}  & 280.5904   & $-2.140$   & 0.000     & $0.58\pm 0.06$ & 1.348 \\
\hline \\[-1.5ex]
Fe\,\textsc{i}  & 279.8600   & $-1.321$   & 7376.764  & $0.64\pm 0.06$ & 0.713 \\
Ni\,\textsc{i}  & 279.9474   & $-1.470$   &  879.813  & $0.68\pm 0.07$ & 1.448 \\
Fe\,\textsc{i}  & 280.5346   & $-0.953$   & 7376.764  & $0.68\pm 0.09$ & 1.175 \\
\hline \\[-1.5ex]
Mn\,\textsc{i}  & 279.5641   & 0.530      &  0.000    & $0.74\pm 0.09$ & 1.892 \\
Fe\,\textsc{i}  & 279.9972   & $-1.373$   & 7728.059  & $0.76\pm 0.07$ & 0.915 \\
Fe\,\textsc{i}  & 281.4116   & $-0.350$   & 7376.764  & $0.76\pm 0.11$ & 1.159 \\
\hline \\[-1.5ex]
Fe\,\textsc{i}  & 278.8927   & $-0.020$   & 6928.268  & $0.88\pm 0.11$ & 1.466 \\
Mn\,\textsc{i}  & 279.9094   & 0.400      & 0.000     & $0.84\pm 0.11$ & 1.430 \\
Mn\,\textsc{i}  & 280.1907   & 0.240      & 0.000     & $0.83\pm 0.10$ & 1.286
\enddata
\tablecomments{$z$ is the approximate height above $\tau_{500}$=1 sampled by the line shifts, and $\sigma(\Delta v)$ is the standard deviation of the velocity shifts of each line for snapshot 385. Line groups of similar formation heights are separated by horizontal lines.}
\end{deluxetable}

\begin{figure*}
\begin{center}
\includegraphics[scale=0.85]{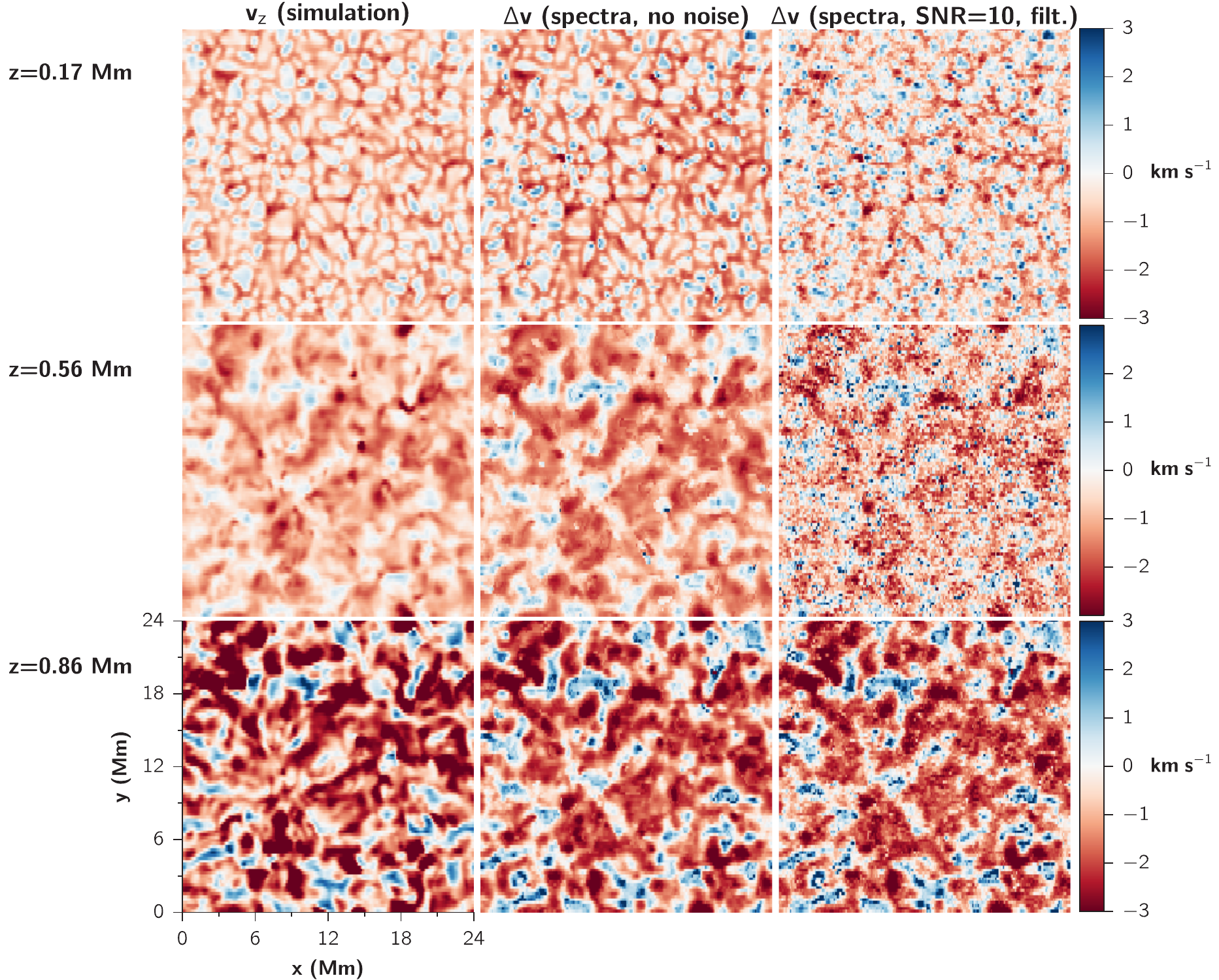}
\end{center}
\caption{Velocity maps for three photospheric heights, showing the atmospheric vertical velocity $v_z$ at a given geometrical height, and the mean velocity shifts $\Delta v$ derived from several photospheric lines. The color scale is clipped at $-3\;\kms$ and $3\;\kms$ to maximize contrast. The middle panels shows the $\Delta v$ when no noise was applied on the spectra, and the right panels for when Poisson noise was added with a S/N of 10, and a Wiener filter was used on the data (see text). \label{fig:vz_panel}}
\end{figure*}

The list of chosen lines and their properties can be found in Table~\ref{tab:lines}. This list is not exhaustive. There are probably other lines in the region that could be useful. However, for our purposes this list forms a reasonable set of lines that are useful for extracting velocities from the synthetic spectra. It is also possible that there are other lines that we have not included in our synthetic spectrum. To keep the computational costs manageable, only the strongest lines were selected. Figure~\ref{fig:mspec_lines} shows that, at least for disk-center, our synthetic spectra reproduce most observed lines reasonably. However, it is possible that some neglected blends will be visible in other solar regions or viewing angles, and potentially interfere with the velocity diagnostics of some of our chosen lines.

Besides velocities from photospheric lines, we also identify `quasi-continuum' regions with few lines blended in the wings of \MgIIhk. These regions are formed much deeper than the cores of \MgIIhk and sample the photosphere; their intensities or radiation temperatures can be used as proxies for the photospheric temperature at different heights. Using the \MgIIhk\ wings to derive temperatures was first done by \citet{Morrison:1978} and similar approaches have been taken for temperatures from the \CaII\ lines \citep[e.g.][]{Sheminova:2005, Henriques:2012, Beck:2013}. Looking at the intensity contribution functions as a function of wavelength, we identified three groups of these quasi-continuum regions whose intensities provide temperature diagnostics at different heights. These are listed in Table~\ref{tab:cont}. The groups are also shown as color shades in Figure~\ref{fig:mspec_lines}.

\begin{deluxetable}{lrrrr}
\tablecaption{Selected Quasi-continuum Regions for Temperature Estimation\label{tab:cont}}
\tablehead{
\colhead{Region} & \colhead{$\lambda_i$} & \colhead{$\lambda_f$} & \colhead{$z$} & \colhead{Group} \\
     & \colhead{(nm)} & \colhead{(nm)} & \colhead{(Mm)}} & 
\startdata
\MgIIk\ far blue wing    & $278.814$  & $278.834$  & $0.15\pm 0.02$ & 1  \\
\MgIIk\ blue wing        & $279.510$  & $279.533$  & $0.42\pm 0.03$ & 3  \\
Bump between \hk         & $280.034$  & $280.051$  & $0.28\pm 0.02$ & 2  \\
\MgIIh\ blue wing        & $280.260$  & $280.283$  & $0.42\pm 0.03$ & 3  \\
\MgIIh\ far red wing     & $281.028$  & $281.047$  & $0.15\pm 0.02$ & 1  
\enddata
\tablecomments{$z$ is the approximate height above $\tau_{500}$=1 sampled by the radiation temperatures; $\lambda_i$ and $\lambda_f$ are respectively the starting and ending wavelengths for each region.}
\end{deluxetable}

\subsection{Extracting Spectral Properties}

The velocity shifts for each line in Table~\ref{tab:lines} were extracted with an automated procedure. For each spectrum, the wavelength shift $\Delta \lambda$ was estimated by taking the centroid of the signal $s(\lambda)$:
\begin{equation}
\Delta\lambda = \lambda_0 - \frac{\int \lambda \cdot s(\lambda) \mathrm{d}\lambda }{\int s(\lambda) \mathrm{d}\lambda},
\end{equation}
where $\lambda_0$ is the line center wavelength. The integrations are performed in a small interval (typically around $\pm$3 pixels) around $\lambda_g$, a guess for the wavelength shift, and in that interval $s(\lambda)$ is defined as:
\begin{equation}
s(\lambda) = 1 - I(\lambda)/I(0),
\end{equation}
where $I(\lambda)$ is the spectral intensity and $I(0)$ its value at the first point of the interval. Here $\lambda_g$ is obtained by taking the minimum value of a local mean spectrum (average spectrogram in a region of about 1\arcsec) near the transition wavelength. The wavelength shift is then converted to velocity shift:
\begin{equation}
\Delta v = \frac{\Delta\lambda}{\lambda_0}c.
\end{equation}
This procedure is repeated for each line and spatial point. This centroid approach was preferred over just taking the minimum of the spectra to allow for sub-pixel precision, and is analogous to the center-of-gravity method described by \citet{Uitenbroek:2003}. As in the center-of-gravity  approach, our method is independent of spectral resolution and not too sensitive to noise.

The photospheric temperature estimates are obtained for each region in Table~\ref{tab:cont}, simply by averaging the spectra in each wavelength window. The intensities are then converted to radiation temperatures using the Planck function, and in turn averaged for the different groups. To improve the statistics, we use the average of two wavelength regions that sample approximately the same formation heights for groups 1 and 3.

\subsection{Velocities\label{sec:ph_vel}}

The velocity shifts extracted from the photospheric lines can be correlated with the atmospheric vertical velocity $v_z$ at a given fixed height $z$. This height is defined as the geometrical height above $z(\tau_{500}=1)$, the horizontal mean height where the optical depth reaches unity at 500~nm. For each line, we identified the approximate atmospheric height that the line shifts sample by finding which value of $z$ minimized the squared difference between the velocity shifts and $v_z$. According to the height sampled, the lines were grouped into eight different bins (see Table~\ref{tab:lines}), and their velocity shifts were averaged to improve the statistics. These values were then compared with $v_z$ at the height closest to the mean $z$ of each group: 0.17, 0.24, 0.38, 0.42, 0.56, 0.68, 0.76, and 0.86~Mm. For each group we tried to obtain as many clean lines as possible, but for some (e.g. $z=0.42$~Mm) this was not possible.

Because the lines are formed in a range of formation heights following a corrugated surface, comparing velocity shifts with velocities at a fixed height is an approximation; the uncertainty of each line's formation height is also noted in Table~\ref{tab:lines}. The uncertainties show that for some lines the different height groupings overlap, illustrating the difficulty in obtaining velocities at a precise height.

In Figure~\ref{fig:vz_panel}, we compare the atmospheric  vertical velocities $v_z$ with velocity maps extracted from the photospheric lines, for snapshot 385, and for three heights: 0.17, 0.56, and 0.86~Mm. The $v_z$ values for each height were spatially convolved in the same manner as the spectra. The velocity shifts can reproduce, to a reasonable degree, the atmospheric velocity structure. The differences are larger for more extreme values; in particular, the predicted upflows tend to be larger by about $1-2\:\kms$. However, given the spectral resolution and pixel sizes of IRIS, the results are very encouraging.

\begin{figure}
\begin{center}
\includegraphics[scale=0.84]{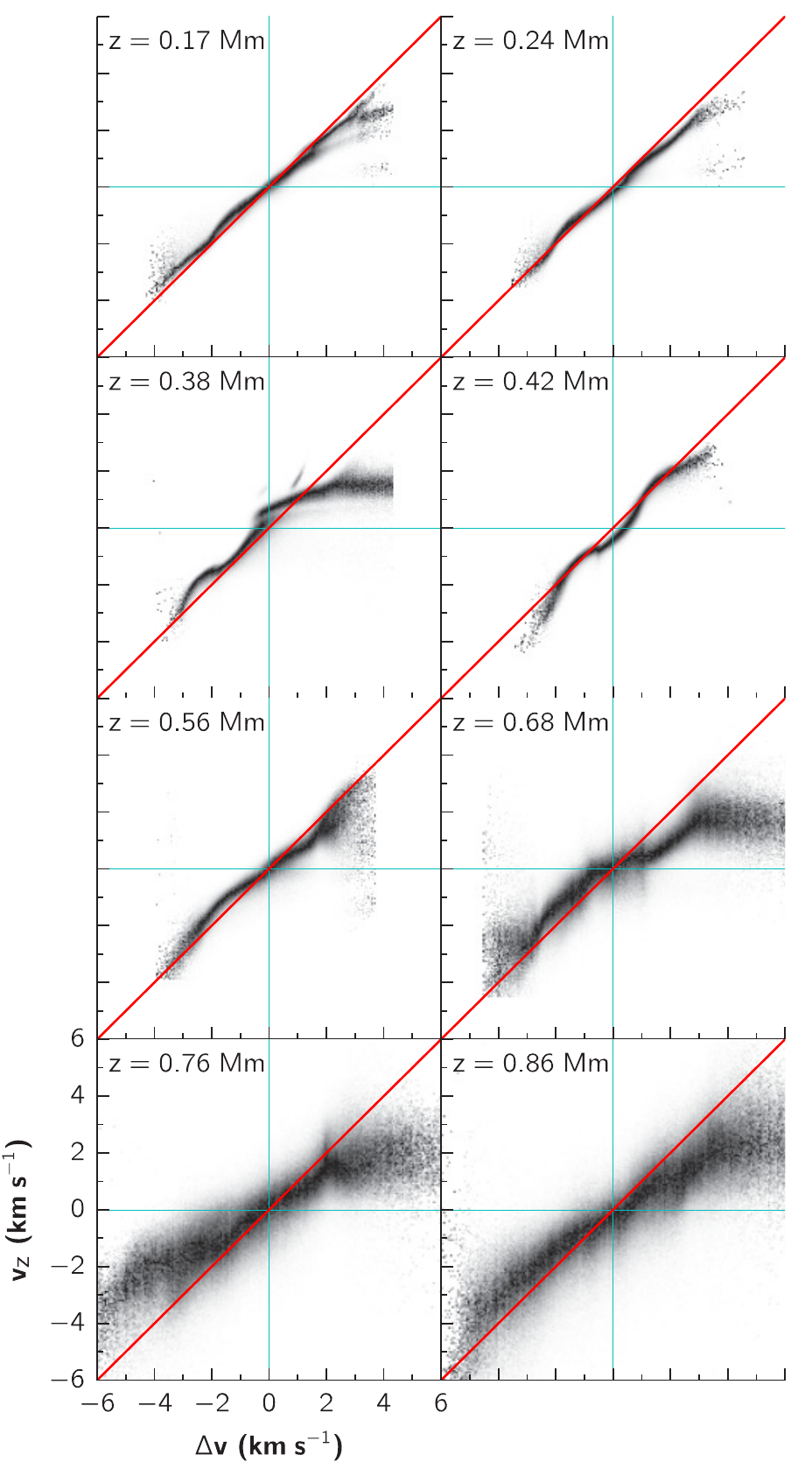}
\end{center}
\caption{PDFs for the atmospheric vertical velocity being $v_z$ when the observed velocity shifts derived from photospheric lines are $\Delta v$, at different depths and combining all snapshots. Darker means more frequent. The diagonal red lines depict $y=x$; the cyan lines depict the zero velocities.\label{fig:ph_vz}}
\end{figure}

\begin{figure*}
\begin{center} 
\includegraphics[scale=0.84]{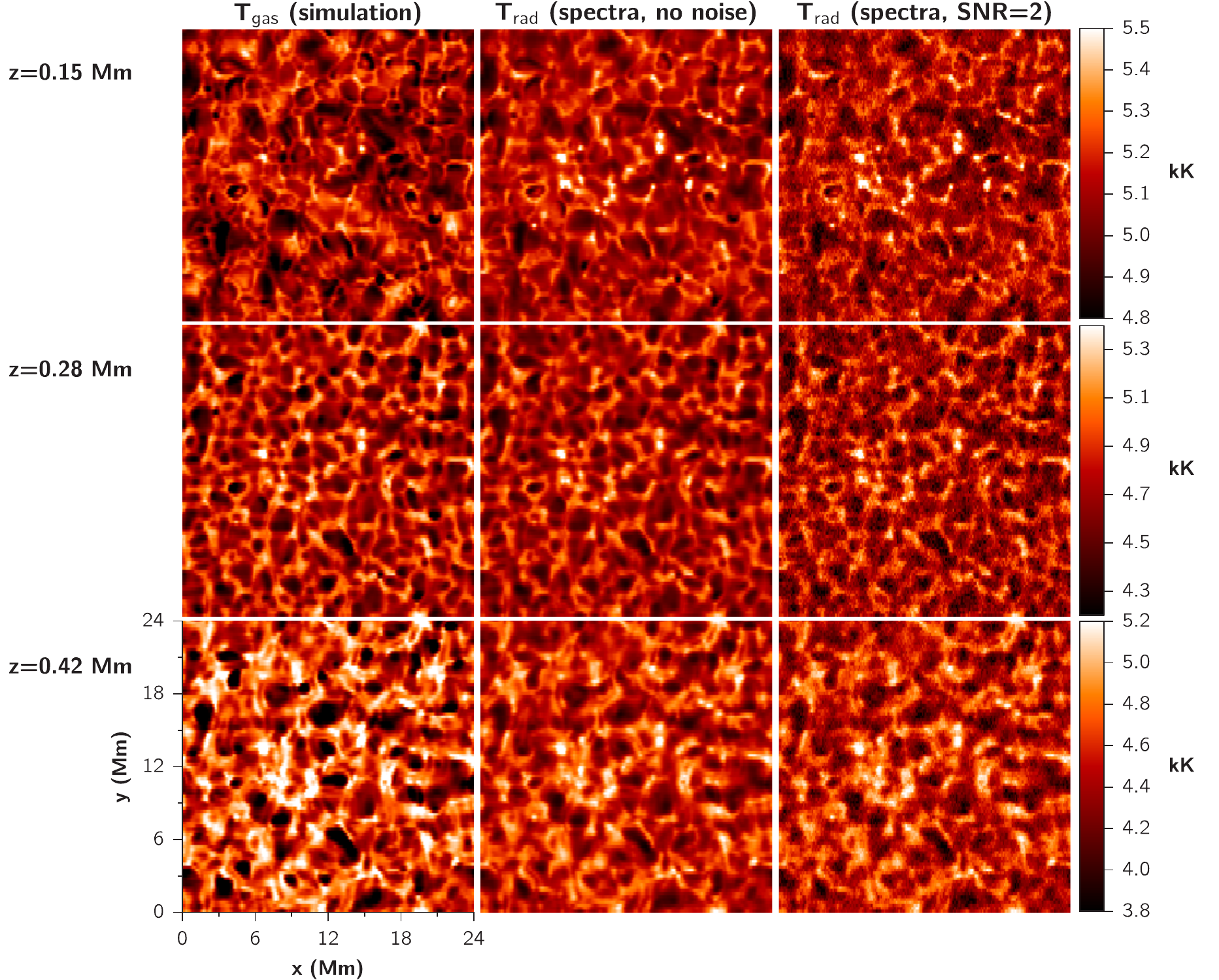}
\end{center}
\caption{Temperature maps for three photospheric heights, showing the atmospheric gas temperature $T_\mathrm{gas}$  at a given geometrical height, and the radiation temperature $T_\mathrm{gas}$ derived from the spectra. The color scale is clipped to maximize contrast. The middle panels shows $T_\mathrm{gas}$ when no noise was applied to the spectra, and the right panels show the same quantity when Poisson noise was added with a S/N of 2.\label{fig:tg_panel}}
\end{figure*}

We combine the results for the 37 snapshots and all heights tested in Figure~\ref{fig:ph_vz}, where we show the scaled PDFs of the relation between the velocity shifts $\Delta v$ (averaged for all the lines sampling a given height) and the spatially convolved atmospheric velocities at each height. As for the Mg diagnostics, these PDFs have been scaled by the maximum value in each column in the x axes. Absolute velocity shifts larger than 6~$\kms$ are not reliable and are not shown. Overall, the agreement is very good. %

As seen in Figure~\ref{fig:ph_vz}, the range of photospheric line shifts goes from at least $-4$ to 4$\;\kms$. In particular for the lines formed deeper, this is larger than what is typically observed in the quiet Sun \citep[e.g.][]{Kiselman:1994, Pereira2009b, Beck:2009}. The reason why such larger velocities occur is related to oscillations in the simulation. These oscillations are stochastically excited by the convective motions in the simulations in much the same way as in the Sun. They are reflected by a pressure node in the bottom boundary of the simulation box rather than being refracted back up as happens in the Sun. Because of the limited size of the box, there is a limited number of modes possible for the oscillations; with a similar total energy, as in the solar case, the amplitude of the modes (dominated by the global mode) is much larger \citep{Stein:2001}. This means that for some snapshots the oscillatory component of the velocity is larger than the convective component, and larger-than-observed line shifts are seen. This higher velocity range does not change our conclusions and in fact provides a more extensive test of our extraction algorithm by sampling a larger range of velocities.

The quality of the velocity estimates varies for different heights. This has several causes. The main factor is the number of lines used to estimate the velocity at each height. Relying on only a few lines makes the result more prone to statistical fluctuations, misidentifications of the automatic procedure, etc. Also, some lines will be more susceptible to the influence of blends, in particular for the more extreme velocity values where a nearby blend could be confused with the line itself. Finally, as one probes higher atmospheric regions, the line radiation has a contribution function that is increasingly corrugated in height and the correlation with a fixed geometrical height starts to break down. 

For the majority of the points, the error in the velocity estimates from the photospheric lines is less than 1~$\kms$. %

\subsection{Temperatures\label{sec:ph_temp}}

To estimate the temperatures at different layers of the photosphere we make use of five wavelength regions, organized in three groups of formation heights (0.15, 0.28, and 0.42~Mm). These regions, shown in Figure~\ref{fig:mspec_lines} and listed on Table~\ref{tab:cont}, probe the deepest photospheric layers sampled by the intensities in the wings of \MgIIhk.

\begin{figure}
\begin{center}
\includegraphics[scale=0.87]{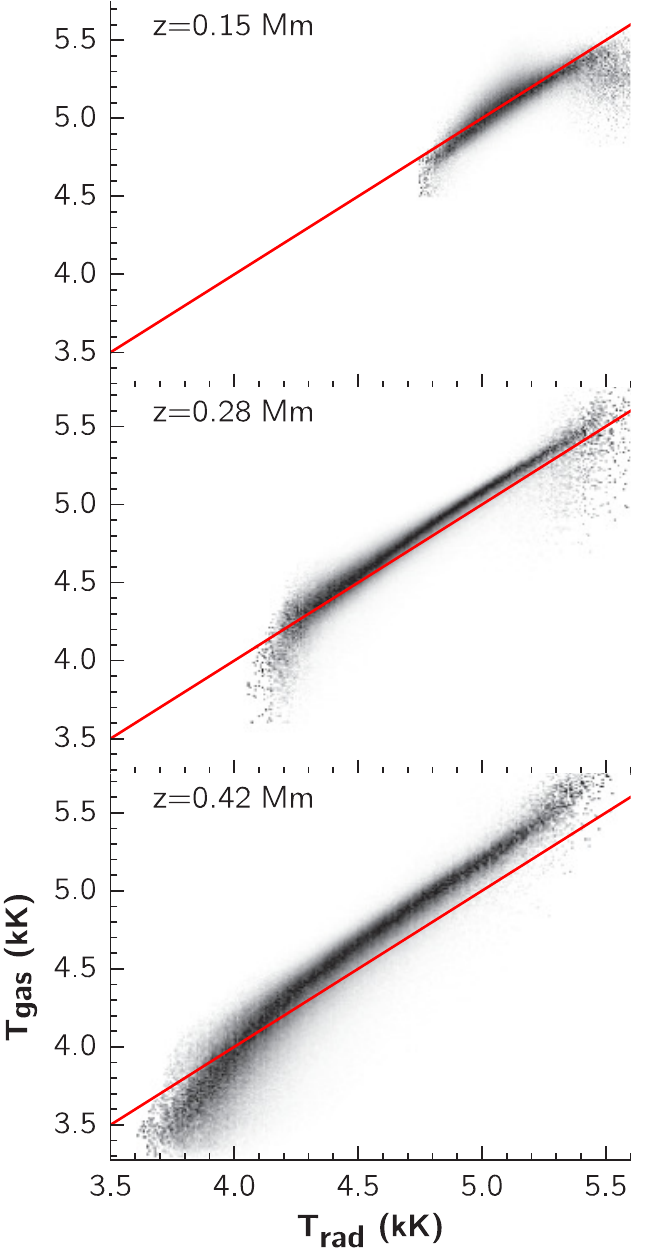}
\end{center}
\caption{Probability density functions for the atmospheric temperature being $T_{\mathrm{gas}}$ when the observed radiation temperature is $T_{\mathrm{rad}}$, at different depths and combining all snapshots. Darker means more frequent. The diagonal red lines depict $y=x$.\label{fig:tg_scatter}}
\end{figure}

In Figure~\ref{fig:tg_panel}, we compare the atmospheric temperatures $T_\mathrm{gas}$ with maps of the radiation temperatures $T_\mathrm{rad}$ extracted from the spectra, for snapshot 385. The $T_\mathrm{gas}$ values for each height were spatially convolved in the same manner as the spectra. It can be seen that the $T_\mathrm{rad}$ values provide a good estimate for the $T_\mathrm{gas}$ values at the different heights. The largest differences between $T_\mathrm{gas}$ and $T_\mathrm{rad}$ occur at the high and low range of temperatures. The estimates from  $T_\mathrm{rad}$ tend to be slightly higher than $T_\mathrm{gas}$, in particular for the very cool pockets and the hottest points.

The relation between spectral intensities and atmospheric temperatures can be better observed in Figure~\ref{fig:tg_scatter}, where we show the scaled PDFs comprising results for all snapshots. One can see that $T_\mathrm{rad}$ is a good estimator for $T_\mathrm{gas}$, with the largest differences seen at the lower temperatures. The best correlation is found for $z=0.28$~Mm, with $\left|T_\mathrm{gas} - T_\mathrm{rad}\right|$ being less than 100~K for most points. At $z=0.42$~Mm, there is a slight systematic shift where  $T_\mathrm{gas}$ is larger by about 200~K. The cool pockets, as shown in Figure~\ref{fig:tg_panel}, are not as pronounced in the $T_\mathrm{rad}$ maps. The main reason for the discrepancy at low $T_\mathrm{gas}$ comes from the fact that the radiation is emitted in a range of heights and very low temperatures at a given geometrical height can be compensated by a slightly warmer layer at a different location along the ray path. Overall, intensities from the quasi-continuum regions provide an excellent temperature diagnostic, with tight linear correlations with the photospheric temperatures at different heights.

\begin{figure}
\begin{center}
\includegraphics[scale=0.87]{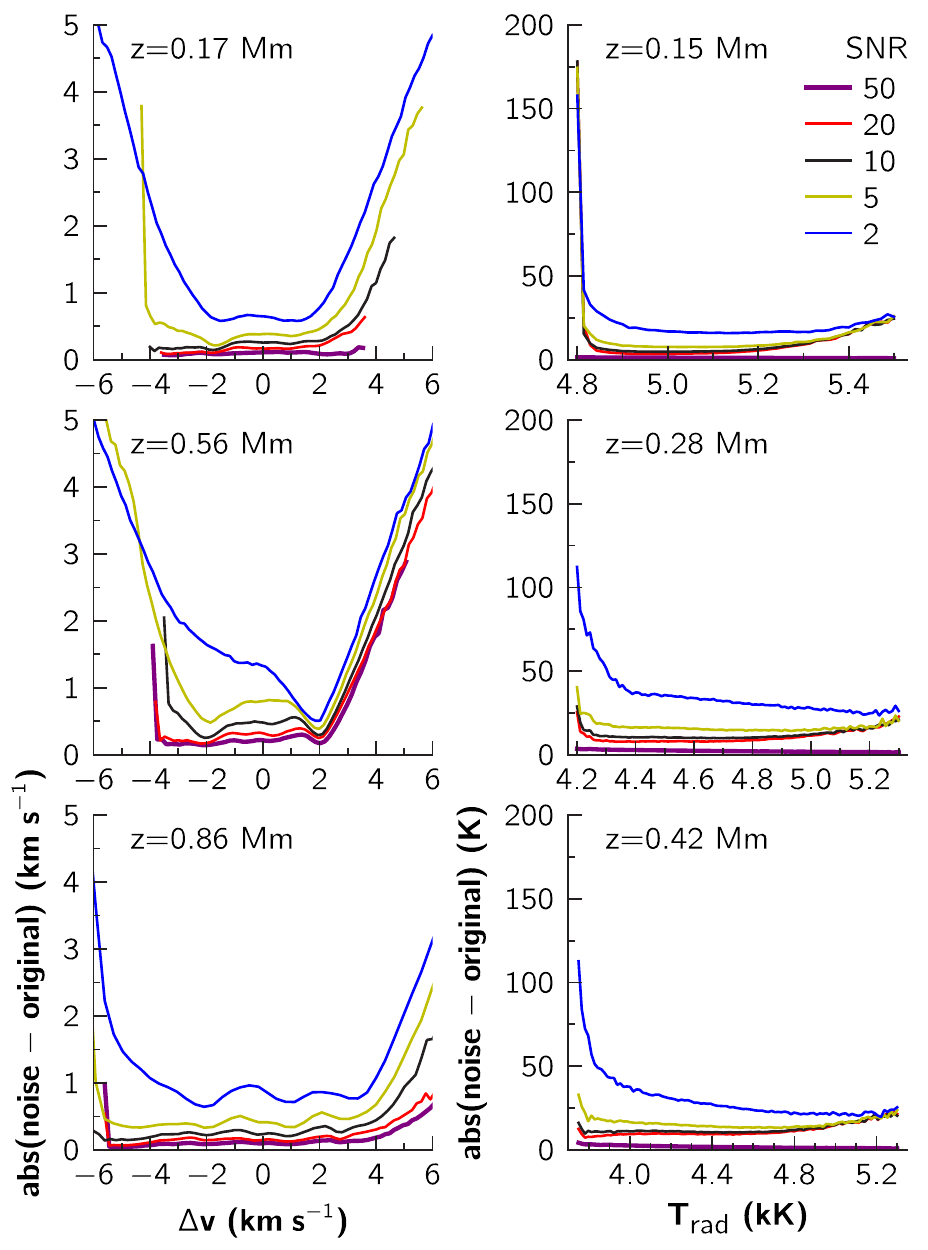}
\end{center}
\caption{Median values of the absolute difference between quantities derived from spectra with and without noise. We show results for different S/Ns, as per the legend. For all S/Ns except 50, a Wiener filter was applied on the spectra to mitigate the effects of noise. \emph{Left panels:} effects of noise on the derived velocity shifts; \emph{right panels:} effects of noise on the derived radiation temperatures.\label{fig:ph_noise}}
\end{figure}

\subsection{Noise Mitigation}

To investigate the effects of noise on the extracted velocities and temperatures, we ran the same procedure on the spectrograms with varying levels of noise. To mitigate the effects of noise, we apply the same Wiener filter as described in Section~\ref{sec:mg_noise}. The results are summarized in Figure~\ref{fig:ph_noise}. As in Figure~\ref{fig:mg_noise} the different lines show the location of the median of the absolute difference between the quantities derived from the spectra with and without noise. The results in Figure~\ref{fig:ph_noise} are from the quantities derived from the Wiener-filtered noisy spectra, except for a S/N of 50 where no filtering was necessary. 

The effects of noise and filtering in the velocities can be also seen in the right panels of Figure~\ref{fig:vz_panel}. A S/N of 10 entails a moderate amount of noise, but one can see that when filtered, the velocity estimates are still reasonable.

The effects of noise are more noticeable for the velocities at $z=0.56$~Mm, most likely because these lines are either weak or in crowded spectral regions. Additionally, by being in a lower mean intensity region (close to the \hk\ peaks), these lines also suffer from a reduced S/N when compared with other lines. In Figure~\ref{fig:ph_noise}, one can see that for the velocities at $z=0.56$~Mm and a $\Delta v$ above 3~$\kms$, the effects of noise increase sharply. (Above a median difference of 1~$\kms$ the correlation between $\Delta v$ and $v_z$ becomes very diffuse.) Generally speaking, the velocity shifts determined by the centroid method are robust enough even at low S/Ns, especially when noise filtering is used. We find that a S/N of 5 is the minimum to obtain velocity estimates that are correlated with the photospheric velocities. The filtering helps even at a S/N as high as 20, but above that level it brings no advantage. 

The photospheric temperature diagnostics are essentially immune to noise. Because the derived $T_\mathrm{rad}$ values are simply obtained by averaging several regions of a spectrum, the noise is averaged out and its effects are only barely noticeable when the intensity is very low. As seen in the right panels of Figure~\ref{fig:ph_noise}, even with the rather extreme S/N of 2 and no filtering, the effect on the temperature maps is small. From  Figure~\ref{fig:ph_noise} one can see that, except for the very low $T_\mathrm{rad}$, the median difference between noise and no noise is less than 50~K, and much less for higher S/Ns. At S/Ns of 2 or 5, as noted for the \MgII\ \ktwoV\ temperatures, the Wiener filter helps cleaning out the quantization introduced by the noise. But for higher S/Ns, a Wiener filter is not usually needed for the temperature estimation.

\section{Slit-jaw images} \label{sec:slit-jaw}

\begin{figure*}
\begin{center}
\includegraphics[scale=0.86]{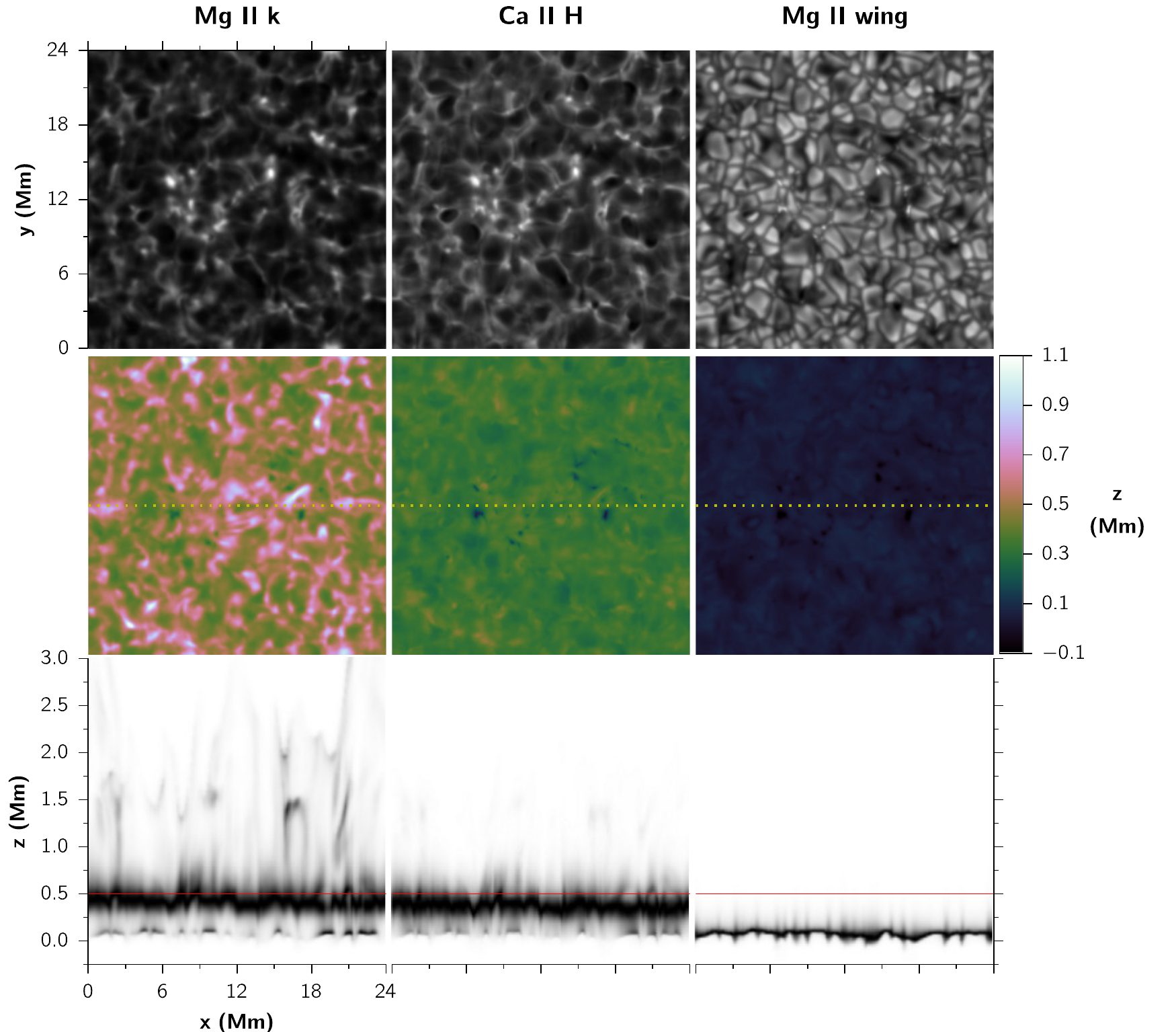}
\end{center}
\caption{Formation of slit-jaw images in snapshot 385. Each column corresponds to a different filter, labeled on top. \emph{Top panels:} simulated slit-jaw images. Intensity scaling was made from the minimum value to 90\% of the maximum value, except for the \MgII\ wing filter where the maximum was used. \emph{Middle panels:} centroid height of the contribution function to intensity (see text). Dotted lines depict $y=12$~Mm, the slice used to calculate the bottom panels. \emph{Bottom panels:} contribution functions to intensity for a horizontal slice at $y=12$~Mm, normalized to the maximum for each point in the $x$ axis. The red line depicts $y=0.5$~Mm and the tickmarks are the same for each panel.\label{fig:sj}}
\end{figure*}

The slit-jaw filtergram images for the NUV window of IRIS provide invaluable context information but also contain diagnostic information by themselves. We investigated the formation properties of the two NUV slit-jaw images, for the \MgIIk\ core and in the far wing of the \MgII\ lines. The former serves as a mid-chromospheric diagnostic and the latter is useful for photospheric context imaging and diagnostics.

When considering the simulated \MgIIk\ filtergrams one should keep in mind the limitations of the simulation. As is visible in Figure~\ref{fig:mspec_lines}, the mean synthetic spectrum exhibits weaker and narrower \MgIIhk\ cores than observed. This limitation, also discussed in \citetalias{Leenaarts:Mg2}, causes our sample of spectra and slit-jaw images to not cover the whole range of conditions observed in the Sun. A stronger \MgIIk\ emission peak will contribute proportionally more intensity in the filtergram. This will cause the images to sample structures formed higher in the chromosphere and limit the relative photospheric contribution into the filtergram. Therefore, the results presented here should be viewed as an example of very quiet Sun and not representative of the many regions of higher activity that IRIS will observe.

In Figure~\ref{fig:sj} we show simulated slit-jaw filtergrams and an overview of their formation properties, for snapshot 385. In the top panels we compare the \MgII\ slit-jaw filtergrams with a \CaH\ filtergram simulated for \emph{Hinode} SOT/BFI. To quantify the formation properties of each filter, we started by calculating the contribution function for the vertically emergent intensity, $C_{I}$, defined as
\begin{equation}
C_{I} = \chi_\nu\;S_\nu\;\mathrm{e}^{-\tau_\nu},
\end{equation}
where $\chi_\nu$ is the total opacity, $S_\nu$ the total source function, and $\tau_\nu$ the optical depth; all of these quantities are functions of wavelength and height. After calculating $C_{I}$ for all spatial points and wavelengths, we applied the different filter transmission properties and obtained the approximate contribution function for each spatial point and filter. Furthermore, for each depth we applied the IRIS spatial convolution and slit-jaw pixel size of $0\farcs166^2$ (or the BFI PSF and $0\farcs0.541^2$ pixels for \CaH). In the middle panels of Figure~\ref{fig:sj} we show the centroid height of $C_{I}$, a proxy for the filter intensity formation height at each column in the simulation, and calculated as
\begin{equation}
C_{Ih} = \frac{\int z \cdot C_{I}(z) \mathrm{d}z}{\int  C_{I}(z)  \mathrm{d}z}.
\end{equation}
In the bottom panels of Figure~\ref{fig:sj}, we show the normalized $C_{I}$ for all spatial points along a horizontal cut at $y=12$~Mm (shown with a dotted line in the middle panels). These  $C_{I}$ values were normalized by their maximum for each column along the $x$ axes. The zero point of the height scale is defined where $\tau_{500}$ reaches unity.

In the solar atmosphere Mg is about 18 times more abundant than Ca \citep{Asplund2009}. All other things being equal, a Mg line will always be formed higher than a Ca line. The oscillator strength of \MgIIk\ is only slightly larger than that of \CaH, so most of the difference in their formation heights will come from the abundance difference. When applying a filter, contribution functions at different wavelengths are weighted by the filter transmission function and the filter width and line width influence the formation height.  In this quiet Sun simulation, the contribution functions of the \MgIIk\ and \CaH\ filtergrams both usually peak at a height around 0.4~Mm, extending down to almost 0~Mm. However, for \CaH\ filtergrams, very little radiation comes from layers above 0.6~Mm, while for \MgIIk\ the contribution function continues to much larger heights, in some cases above 3~Mm. This stark difference is evident in the figure of the centroid heights of $C_{I}$, showing that for the regions where there is more chromospheric activity this is picked up in \MgIIk\ but not in \CaH.  The \CaH\ line profile is narrower than \MgIIk, with the higher-intensity wings closer to the line core. Even when its core is in emission and taking into account the narrower BFI filter transmission profile, it has a larger contribution of the wings in the filtergram intensity. On the other hand, the far wings of \MgIIk\ are wider and the emission core rises out of a much darker vicinity (further helped by the PRD in the line core), which increases the contribution of upper chromospheric light.

The formation properties of the BFI \CaH\ filter have been studied before by \cite{Carlsson:2007} and \cite{Beck:2013b}. Both of these studies find that the \CaH\ filter is formed at a lower height ($\approx0.2$~Mm) than the 0.4~Mm we find here. However, both studies used the response function, a different measure from the $C_{I}$ we used here, and both of them used 1D semi-empirical atmospheres, whereas we employ a 3D simulation.

When compared with \CaH, the \MgIIk\ filtergrams have a darker background. The $\Delta I/\langle I \rangle$ of the \MgIIk\ filtergrams is 53\%, while for the \CaH\ filtergrams it is 27\%. This difference makes it easier to identify bright dynamical features in \MgIIk, because the background is much darker. Most of this is a consequence of its spectral profile having an emission peak and low-intensity wings, reflecting its larger formation height. Some of the contrast difference can also be explained by the higher sensitivity of the Planck function at shorter wavelengths. For this particular simulation, the structures probed in both filtergrams appear similar, which may not always be the case. Taking into account the much higher \MgIIk\ peak emission observed in the Sun, we expect the chromospheric contribution to the \MgIIk\ filtergrams to be even larger, leading to different structures probed when comparing with the \CaH\ filtergrams.

The \MgII\ wing continuum filtergram is mostly photospheric in origin. It is formed just above 0~Mm, in a much narrower region when compared with the other two filtergrams. Its radiation temperature has a good correlation with the gas temperature at a fixed height. We find that the best correspondence is found at $z\approx0.07$~Mm.

\section{Discussion} \label{sec:discussion}

\subsection{Limitations and Comparison with Observations}

Using a realistic simulation of the quiet Sun and detailed radiative transfer calculations, we studied several diagnostics provided by the NUV window of IRIS and how they are affected by the instrumental resolution and noise. One of the main caveats of this approach is the limitations of the simulation employed here, in particular regarding the \MgIIhk\ diagnostics. These limitations, discussed in more detail in \citetalias{Leenaarts:Mg2}, lead to simulated \MgIIhk\ emission profiles that are weaker and narrower than what is observed (see Figure~\ref{fig:mspec_lines}). These weaker profiles will limit our sample of synthetic spectra. While we believe that our sample is extensive enough to cover a good amount of quiet Sun conditions, an important question to ask is how the derived relationships would change under more realistic conditions, and for different solar regions (e.g. active regions). Preliminary results on the \MgII\ diagnostics using different simulations with stronger magnetic fields and more dynamic chromospheres indicate that most of the relationships still hold, in particular the \kthree/\hthree\ velocities and \ktwo/\htwo\ peak intensity diagnostics. However, a definite answer will require more extensive modeling and a better understanding of the chromosphere. The emission from the \hk\ lines is greatly increased in active regions \citep[see][]{Morrill:2008}, so one should be careful to not overinterpret our results when comparing with observations in these very different conditions.

The radiation of \kthree\ and \hthree\ is affected by 3D effects, because for these features the photon mean free paths are longer than the typical horizontal structure sizes \citepalias[see][]{Leenaarts:Mg1}. Because the computational costs would be too high otherwise, we studied these features using synthetic spectra calculated using the 1.5D approximation (unlike in \citetalias{Leenaarts:Mg2}, where 3D CRD calculations were used for one snapshot). However, the 3D effects in these features are seen mostly in the intensities and formation heights, and are not as extreme for velocity shifts (even at the most extreme cases, differences between 1.5D and 3D shifts are seldom more than 2~$\kms$). In particular, the relation between $v_z(\tau=1)$ and the velocity shifts still holds when using the 1.5D approximation, even if the formation heights are not the same as in 3D. Thus, we believe that using 1.5D for these velocity shifts is justified; the effects of noise and the finite resolution of IRIS should apply equally for a full 3D calculation.

Apart from the shortcomings of the \MgIIhk\ emission profiles, our synthetic spectra including the many photospheric lines compare very favourably with the observations in Figure~\ref{fig:mspec_lines}. Most of the lines included match an observed feature, and even when only the lower-resolution observations are shown, the synthetic and observed mean spectra are still very close. This gives us confidence about the realism of the simulated photosphere and the line synthesis. 

\subsection{Summary of \MgIIhk\ diagnostics}

\begin{deluxetable}{ll}
\tablecaption{Correlation between \MgII\ Features and Atmospheric Properties.\label{tab:mgsummary}}
\tablehead{
\colhead{Spectral Observable} & \colhead{Atmospheric Property}}
\startdata
$\Delta v_{\mathrm{k3}}$ or $\Delta v_{\mathrm{h3}}$ & Upper chromospheric velocity  \\
$\Delta v_{\mathrm{k2}}$ or $\Delta v_{\mathrm{h2}}$ & Mid chromospheric velocity  \\
$\Delta v_{\mathrm{k3}} - \Delta v_{\mathrm{h3}}$ & Upper chromospheric velocity gradient \\
k or h peak separation & Mid chromospheric velocity gradient \\
\ktwo\ or \htwo\ peak intensities &  Chromospheric temperature \\
$(I_{\mathrm{k}2v} - I_{\mathrm{k}2r}) / (I_{\mathrm{k}2v} + I_{\mathrm{k}2r})$ & Sign of velocity above $z(\tau=1)$ of \ktwo$\dagger$
\enddata
\tablecomments{This is a simplified view, and all correlations above have scatter. $\dagger$ Likewise for the \htwo\ peaks. }
\end{deluxetable}

The \MgIIhk\ lines are the most promising chromospheric diagnostics in the NUV window of IRIS. They provide robust diagnostics of temperature and velocity at different chromospheric heights. The formation of k and h is very similar, with both lines sharing the lower level. The k line has twice the oscillator strength of the h line.

In Table~\ref{tab:mgsummary} we summarize how different observable quantities translate into atmospheric properties. Note that some of them have significant uncertainties, see below.

The shifts of the \kthree\ and \hthree\ minima provide a robust measure of the velocity at the upper chromosphere. Their correlation with the line-of-sight atmospheric velocity at their $\tau=1$ heights is very tight and the most robust of all the synthetic observables tested. The k and h lines are formed similarly since the lines have the same lower atomic energy level. The k line has an oscillator strength twice that of the h line, leading to a difference between the \kthree\ and \hthree\ $\tau=1$ height of about 50~km \citepalias{Leenaarts:Mg2}. This difference can be explored to measure dynamical phenomena. We find that the difference in velocity shifts between \kthree\ and \hthree\ has a good correlation with the atmospheric velocity difference between those two heights of formation, enabling an observer to measure accelerations in the chromosphere. The \kthree\ and \hthree\ intensities show a weak anti-correlation with their formation height. At the resolution of IRIS, this correlation is less tight, but within small windows (a few Mm or less) the intensities can be used to identify local variations in $\tau=1$ heights. From \citetalias{Leenaarts:Mg2}, we know that \hthree\ is typically formed 200~km below the transition region, so the \kthree\ and \hthree\ intensities can provide a measure of the local corrugation of the transition region height.

The \ktwo\ and \htwo\ peaks provide additional diagnostic information. Their peak intensities are correlated with the atmospheric temperature at their $\tau=1$ heights. For low radiation temperatures (below 6~kK), the relation has a certain amount of scatter, but it improves for higher radiation temperatures. The radiation temperature of the \ktwo\ or \htwo\ peaks provides a very tight constraint on the minimum gas temperature: $T_{\mathrm{gas}}$ is seldom less than $T_{\mathrm{rad}}$. The ratio of the red and blue peak intensities  also provides an important measure of the average velocity between the \kthree/\hthree and the \ktwo/\htwo\ formation heights; strong ratios can be used to identify strong upflows or downflows. Much like for \kthree/\hthree, the mean velocity shifts of the \ktwo\ or \htwo\ peaks are correlated with the atmospheric velocity at their $\tau=1$ height. These peaks are typically formed about 1~Mm below \kthree/\hthree\ \citepalias{Leenaarts:Mg2} and they provide an estimate for atmospheric velocities at these lower chromospheric heights. Additionally, for both lines, the peak separation can be used to estimate the atmospheric velocity gradients in their formation range, although this relation is not very tight in the simulation tested here.

We find that the instrumental resolution has a small effect on the \MgII\ diagnostics. The quantities that only depend on one observable (e.g., the \kthree\ velocity or the radiation temperature of \ktwo) are the least affected  and the effects are larger for those quantities that depend on two observables (e.g., the peak separation or $\Delta v_{\mathrm{k}3} - \Delta v_{\mathrm{h}3}$, the difference in velocity shifts from \kthree\ and \hthree). The \kthree\ and \hthree\ velocities are one of the most reliable diagnostics, suffering little from instrumental resolution and withstanding S/Ns of at least 5 after noise filtering. Because of their differential nature, the peak separation and  $\Delta v_{\mathrm{k}3} - \Delta v_{\mathrm{h}3}$ suffer more from the instrumental resolution, but are still useful velocity diagnostics. The radiation temperatures of the \ktwo\ and \htwo\ peaks, despite having their range narrowed from the instrumental resolution, still very much reflect the same relation with gas temperature as found in \citetalias{Leenaarts:Mg2}. Additionally, they are surprisingly resistant to the effects of noise, being useful even at a S/N as low as 2. Other diagnostics that use intensities, such as the ratio of the peak intensities or the \kthree/\hthree\ intensities, also seem not to be significantly affected by the instrumental resolution (besides the obvious loss in spatial resolution). In summary, the \MgII\ diagnostics still hold when taking the instrumental resolution into account. For S/Ns below 10, noise filtering becomes necessary to get the most out of the data.

\subsection{Summary of Photospheric Diagnostics}
The many photospheric lines present in the NUV window bring complementary information to the \MgII\ lines. Here we focused on two key diagnostics: photospheric velocities and temperatures. Selecting several lines in spectral regions as clean as possible, we could obtain reliable velocity estimates for eight photospheric heights, ranging from 0.17 to 0.86~Mm. Some of these heights are covered by several lines, meaning one can either use several lines to derive a cleaner velocity estimate or only use a few lines when observational programs focus on smaller spectral windows. Generally speaking, the lower the photospheric height the more reliable the velocity estimates. This is related to the complexity of spectral regions but also because of the increased corrugation of formation heights as the lines get stronger. Using blend-free, near-continuum regions in the wings of the \hk\ lines, we also derived estimates for photospheric temperatures at the heights of 0.15, 0.28, and 0.42~Mm. With some limitations at the more extreme ranges, these provide a robust estimate of temperature from photospheric to lower chromospheric heights. For the photospheric velocity estimates we again find it desirable that noise filtering be applied at S/Ns below 10; the temperature estimates suffer very little from even large amounts of noise.

\subsection{Outlook for IRIS and Further Work}

The IRIS mission will provide an unprecedented view of the solar atmosphere, connecting the photosphere, chromosphere and corona. Its NUV window brings exciting new diagnostics that will be available in fast cadences at a spatial resolution never seen before in this region. Using first-principles forward modelling, we showed how the \MgIIhk\ spectra can be used to derive a significant amount of chromospheric information: velocities at different heights, velocity gradients, temperatures, etc. In addition, the many other lines in the region provide additional information from the photosphere. 

The analysis presented here and in \citetalias{Leenaarts:Mg1} and \citetalias{Leenaarts:Mg2} pertains to spectra observed at disk center. \MgIIhk\ diagnostics at the limb present different challenges and will be addressed in a future work.

\begin{acknowledgements}
   T.M.D.P. was supported by the NASA Postdoctoral
   Program at Ames Research Center (grant NNH06CC03B).
   This research was supported by the
   Research Council of Norway through the grant ``Solar Atmospheric
   Modelling'' and through grants of computing time from the Programme
   for Supercomputing, by the European Research Council under the European 
   Union's Seventh Framework Programme (FP7/2007-2013) / ERC Grant 
   agreement No. 291058, and by computing project s1061 from the High End
   Computing Division of NASA. 
   B.D.P. acknowledges support from NASA grants NNX08AH45G, 
   NNX08BA99G, NNX11AN98G, NNM07AA01C (\emph{Hinode}), and NNG09FA40C (\emph{IRIS}). 
   We thank the referee for several useful suggestions that improved the manuscript.
\end{acknowledgements}

\bibliographystyle{apj}
\bibliography{tiago}

\end{document}